\pdfoutput=1
\documentclass[
  10pt,
  conference,
  ]{IEEEtran}

\usepackage[utf8]{inputenc}
\usepackage[T1]{fontenc}
\usepackage[scaled=0.8]{beramono}
\usepackage{listings}
\usepackage{caption}
\usepackage{subcaption}
\usepackage{xcolor}
\usepackage{hyperref}
\usepackage{tikz}
\usepackage{lipsum,adjustbox}
\usepackage{soul}
\usepackage{graphicx}
\usepackage{amsmath}
\usepackage{xspace}
\usepackage{todonotes}
\usepackage{wrapfig}
\usepackage{algorithm2e}
\usepackage[capitalise]{cleveref}
\usepackage{csquotes}
\usepackage{siunitx}
\usepackage{pgfplots}
\usepackage{flushend}

\def\BibTeX{{\rm B\kern-.05em{\sc i\kern-.025em b}\kern-.08em
    T\kern-.1667em\lower.7ex\hbox{E}\kern-.125emX}}

\begin{document}

\newcommand{\Legion}{\textsc{Legion}\xspace}
\newcommand{\AFLNetLegion}{\textsc{AFLNetLegion}\xspace}
\newcommand{\AFLNet}{\textsc{AFLNet}\xspace}
\newcommand{\APPF}{\textsc{APPFuzzing}\xspace}
\newcommand{\NLST}{\texttt{NLST}\xspace}

\newcommand{\UCT}{\mathit{UCT}}

\newcommand{\white}{hollow\xspace}
\newcommand{\red}{solid\xspace}
\newcommand{\black}{redundant\xspace}
\newcommand{\purple}{phantom\xspace}
\newcommand{\golden}{simulation\xspace}
\newcommand{\White}{Hollow\xspace}
\newcommand{\Red}{Solid\xspace}
\newcommand{\Black}{Redundant\xspace}
\newcommand{\Purple}{Phantom\xspace}
\newcommand{\Golden}{Simulation\xspace}

\newcommand{\AFL}{AFL\xspace}
\newcommand{\ProFuzzBench}{\texttt{ProFuzzBench}\xspace}

\newcommand{\aflnetff}{\textsc{AFLNet\_FF}\xspace}
\newcommand{\aflnetbb}{\textsc{AFLNet\_BB}\xspace}
\newcommand{\aflnetrr}{\textsc{AFLNet\_RR}\xspace}
\newcommand{\legionuu}{\textsc{AFLNetLegion\_UU}\xspace}
\newcommand{\legionur}{\textsc{AFLNetLegion\_UR}\xspace}
\newcommand{\legionrr}{\textsc{AFLNetLegion\_RR}\xspace}

\newcommand{\proftpd}{\texttt{ProFTPD}\xspace}
\newcommand{\glob}{\texttt{GLOB}\xspace}
\newcommand{\gcov}{\texttt{Gcov}\xspace}
\newcommand{\random}{\texttt{RANDOM}\xspace}
\newcommand{\roundrobin}{\texttt{ROUND\_ROBIN}\xspace}
\newcommand{\favour}{\texttt{FAVOUR}\xspace}
\newcommand{\mctsr}{\texttt{MCTS\_RANDOM}\xspace}
\newcommand{\mctsp}{\texttt{MCTS\_PARTIAL}\xspace}
\newcommand{\mctsf}{\texttt{MCTS\_FULL}\xspace}

\newcommand{\Python}{\texttt{Python\,3}\xspace}
\newcommand{\C}{\texttt{C}\xspace}
\newcommand{\union}{\cup}
\newcommand{\bparagraph}[1]{\textbf{#1}} %{\noindent\textbf{#1}}
\newcommand{\iparagraph}[1]{\textit{#1}} %{\noindent\textbf{#1}}
\newcommand{\prebparagraph}{} %{\\[-0.8em]}

\newcommand{\FIXME}[1]{\colorbox{yellow}{\bf FIXME: }{\bf #1}}
\newcommand{\FIXED}[1]{\colorbox{green}{\bf FIXED: }{\bf #1}}
\newcommand{\MYTODO}[1]{\colorbox{red}{\bf TODO: }{\bf #1}}

\lstdefinestyle{aflnet}{
  basicstyle=\small\ttfamily,
  frame=single,
  morecomment=[f][\color{gray}][0]{\#},
  % emph={NLST},
  emphstyle=\bfseries,
  basicstyle=\normalsize\ttfamily,
  % otherkeywords={USER, PASS, MKD, STOR},
  % morekeywords={USER, PASS, MKD, STOR},
  % keywordstyle=\bfseries,
}

\lstdefinestyle{customc}{
  breaklines=true,
  frame=none,
  language=C,
  numbers=left,
  showstringspaces=false,
  basicstyle=\scriptsize\ttfamily,
  keywordstyle=\bfseries\color{blue},
  commentstyle=\slshape\color{red!50!black},
  identifierstyle=\color{black},
  stringstyle=\color{orange},
}

\usetikzlibrary{calc,positioning, shapes.geometric, fit, backgrounds, patterns, decorations.pathreplacing}

\usetikzlibrary{calc,positioning, shapes.geometric}
\graphicspath{{Figures/}}
\lstset{inputpath=Listings/}
\makeatletter
\def\input@path{{Figures/}{Sections/}}
\makeatother

% \title{Will state selection algorithms significantly affect network fuzzers' performance?\\
\title{State Selection Algorithms and Their Impact on The Performance of Stateful Network Protocol Fuzzing\\
% should not be used \\\FIXME{A better paper name?\\}}
% \thanks{This research was supported by Data61 under the Defence Science and Technology Group's Next Generation Technologies Program.}
}

% \author{\IEEEauthorblockN{(anonymous authors)}}
\author{\IEEEauthorblockN{
Dongge Liu\IEEEauthorrefmark{1}, Van-Thuan Pham\IEEEauthorrefmark{1},
Gidon Ernst\IEEEauthorrefmark{2},
Toby Murray\IEEEauthorrefmark{1}, and Benjamin I.P. Rubinstein\IEEEauthorrefmark{1}
\\
}
% \IEEEauthorblockA{\IEEEauthorrefmark{1}School of Computing and Information Systems\\
% The University of Melbourne, Melbourne, Australia\\
% Email: donggel@student.unimelb.edu.au,\\
% \{thuan.pham, toby.murray, benjamin.rubinstein\}@unimelb.edu.au\\}
% \IEEEauthorblockA{\IEEEauthorrefmark{2}Software and Computational Systems Lab\\
% LMU Munich, Munich, Germany\\
% Email: gidon.ernst@lmu.de}
% }
\IEEEauthorblockA{\IEEEauthorrefmark{1}
The University of Melbourne, Melbourne, Australia\\
Email: donggel@student.unimelb.edu.au, 
\{thuan.pham, toby.murray, brubinstein\}@unimelb.edu.au\\}
\IEEEauthorblockA{\IEEEauthorrefmark{2}
LMU Munich, Munich, Germany\\
Email: gidon.ernst@lmu.de}
}

\maketitle

\begin{abstract}
  The statefulness property of network protocol implementations poses a unique challenge for testing and verification techniques, including Fuzzing. Stateful fuzzers tackle this challenge by leveraging state models to partition the state space and assist the test generation process. Since not all states are equally important and fuzzing campaigns have time limits, fuzzers need effective state selection algorithms to prioritize progressive states over others. Several state selection algorithms have been proposed but they were implemented and evaluated separately on different platforms, making it hard to achieve conclusive findings. In this work, we evaluate an extensive set of state selection algorithms on the same fuzzing platform that is AFLNet, a state-of-the-art fuzzer for network servers. The algorithm set includes existing ones supported by AFLNet and our novel and principled algorithm called AFLNetLegion. The experimental results on the ProFuzzBench benchmark show that (i) the existing state selection algorithms of AFLNet achieve very similar code coverage, (ii) AFLNetLegion clearly outperforms these algorithms in selected case studies, but (iii) the overall improvement appears insignificant. These are unexpected yet interesting findings. We identify problems and share insights that could open opportunities for future research on this topic.

\end{abstract}


\section{Introduction}
\label{sec: introduction}

Network protocols, such as the Simple Mail Transfer Protocol (SMTP), Real Time Streaming Protocol (RTSP), and Secure Socket Layer Protocol (SSL), are important. They enable consumer communication and recreational services (e.g., email and entertainment services) as well as business and government critical services (e.g., banking and cyber physical systems). %These services make our work more productive and our life more enjoyable so we rely on their security and reliability. 
A single critical vulnerability in these services could lead to catastrophic consequences. In 2001, the Code-Red worm infected more than three hundred thousands computers in less than 14 hours via Hypertext Transfer Protocol (HTTP) queries and caused at least \$2.6 billion in damages globally \cite{codered}. In 2017, the infamous WannaCry ransomeware \cite{wannacry} exploited the Server Message Block (SMB) protocol, causing hundreds of millions of dollars damage. These incidents call for a thorough examination of network protocol implementations, especially network servers as they are usually Internet-facing. Given a server's Internet Protocol (IP) address and its port number, attackers---from anywhere on Earth---can send crafted requests aiming to discover vulnerabilities and develop harmful exploits.

\begin{figure}
  \centering\includegraphics[width=0.35\textwidth]{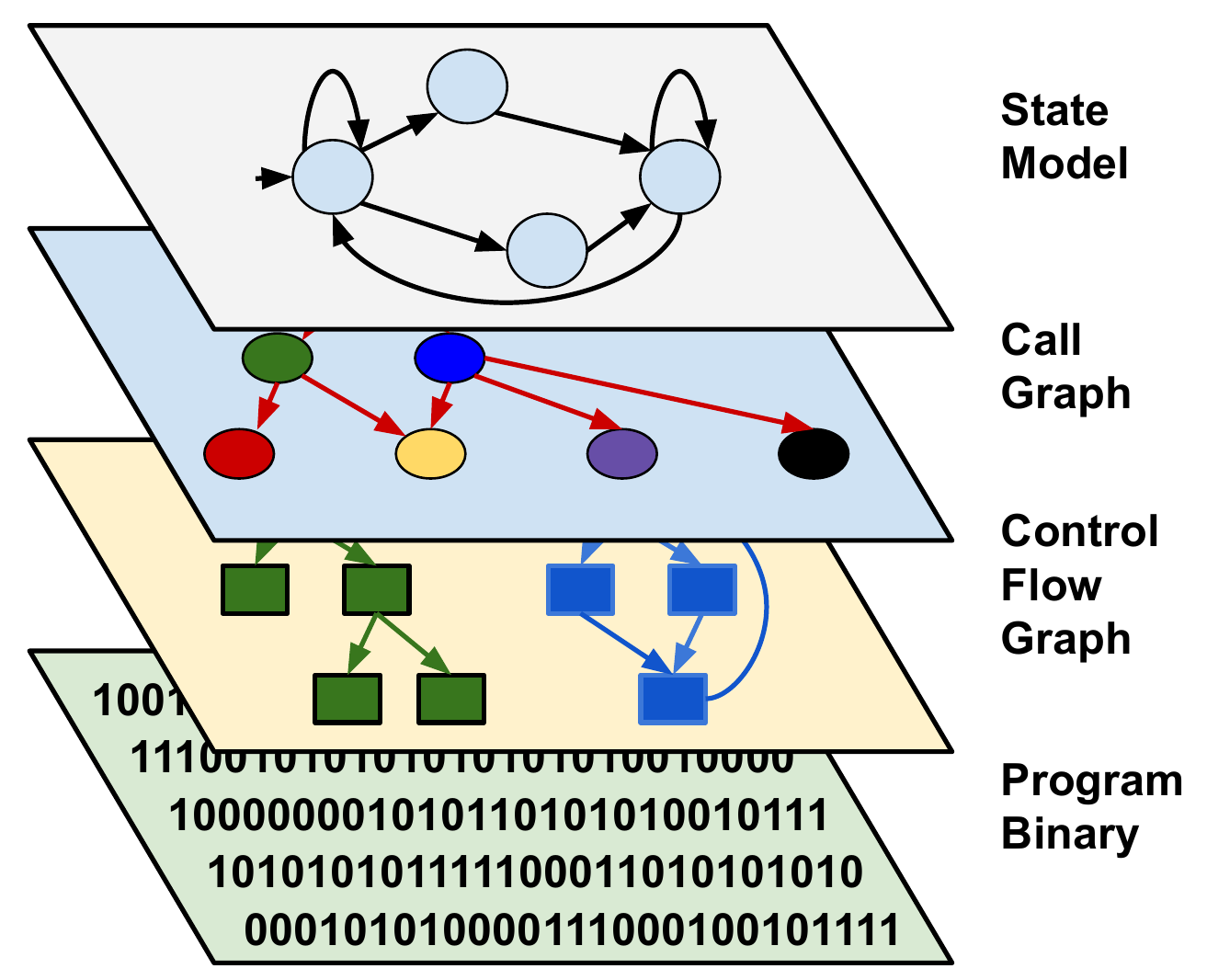}
  %\vspace{-5mm}
  \caption{Four abstraction layers of a server under test.}
  \label{fig:layers}
  %\vspace{-1em}
\end{figure}

Fuzzing (a.k.a Fuzz Testing) \cite{MillerFuzz, ManesSurvey}---as recommended by the U.S. National Institute of Standards and Technology (NIST) \cite{NistGuideline}---is an effective technique for security testing. It has been used to discover thousands of vulnerabilities in large real-world systems \cite{clusterfuzz, onefuzz}. However, fuzzing network servers is challenging because servers are mostly \emph{stateful}. That is, the server under test (SUT) takes sequences of messages (requests sent from the client) and its behaviour depends on both the current message and the current internal server state which is determined by earlier messages. State-of-the-art stateful fuzzers (e.g. Peach \cite{peach}, BooFuzz \cite{boofuzz}, Pulsar \cite{pulsar}, AFLNet \cite{aflnet}, MACE \cite{mace})---including black-box, grey-box, and white-box approaches---tackle that challenge by modelling the SUT's state space and rely on the so-called \emph{state model} to explore the space with \emph{state selection} and test generation algorithms. As depicted in Figure~\ref{fig:layers}, the state model is another abstraction layer of the SUT; it positions on top of the SUT's call graph, control flow graph, and the program binary.

There is a large body of research evaluating exploration algorithms
at the program call graph and control flow graph levels to improve the effectiveness and efficiency of fuzzing \cite{ManesSurvey, aflgo, klee, MoWF,choppedse}. One critical research topic is to study the impact of path, function, and basic block selection algorithms on fuzzing performance. Meanwhile, research on state selection algorithms---that work at the state model layer---is overlooked. 
% \MYTODO{Hi Toby: should we say something about state selection algorithms for model checking etc here. I am not familiar with MC}. 
For instance, in the recent AFLNet paper \cite{aflnet}, only one state selection algorithm was evaluated. Similar is true in many other papers \cite{pulsar, mace, stateafl}. Researchers tend to evaluate their algorithms, as implemented in a whole system, in comparison with other systems e.g. grey-box fuzzing versus black-box fuzzing or greybox-box fuzzing versus white-box fuzzing. In this paper, we evaluate an  extensive set of state selection algorithms on \emph{the same fuzzing platform}.

% While there is a large body of research evaluating state model construction and inference algorithms [cite] as well as test generation algorithms [cite], research on state selection algorithms is overlooked. To the best of our knowledge, there has been no research yet evaluating the impact of state selection algorithms on the performance of stateful network protocol fuzzing. In this work, we aim to fill the gap.

To conduct this research, we need to choose a representative base fuzzer that has existing state selection algorithms and supports an interface to easily implement new ones. Based on this criterion, we chose AFLNet \cite{aflnet}, the first greybox fuzzer for stateful network protocol fuzzing, and built on top of American Fuzzy Lop (AFL) \cite{afl}, one of the most popular and effective fuzzers today. Other fuzzers we considered do not support dynamic state selection algorithms \cite{peach, boofuzz}, or their source code is not publicly available \cite{mace}. 

% \MYTODO{Add justifications for not using PULSAR, PRISMA, and fuzzowski}

We divided our investigation into two phases. In the first phase, we conducted a preliminary study of  three existing state selection algorithms of AFLNet. Surprisingly, the results of 24-hour experiments on six subjects in the ProFuzzBench benchmark \cite{profuzzbench} show that the three algorithms achieved very similar results in terms of code coverage. 
% \MYTODO{Dongge: Show how similar they are quantitatively e.g., differences are less then XXX percent. (1. Run a t-test on the final coverage scores, that p-value goes to here; 2. Compare their similarity across the whole 24h, that goes to Sec. 3)} 
Specifically, the maximum difference between average branch coverage of all benchmarks is $1.15\%$.
% This conclusion is also confirmed by a student t-test. 
Our deep dive into the implementation of AFLNet and those algorithms revealed problems leading to those unexpected results. First, AFLNet's state machine is too coarse-grained resulting in unreliable estimation of each state's potential. Second, its existing algorithms lack a principled way to balance the exploration-exploitation trade-off.
% \MYTODO{Thuan: Problem 1: "compact" state machine, Problem 2: formulas to calculate the scores}.

%--- Old version ---
% We divided our investigation into two phases. In the first phase, we conducted a preliminary study of three existing state selection algorithms of AFLNet. The algorithms are RANDOM (i.e., states are selected randomly), ROUND-ROBIN (i.e., states are stored in a kind-of circular queue), and FAVOR (i.e., states are selected based on some run-time scores). Since FAVOR is the default state selection algorithm of AFLNet, we assumed it would be the best performer. Surprisingly, the results of 24-hour experiments on six subjects in the ProFuzzBench benchmark \cite{profuzzbench} show that the three algorithms achieved very similar results in terms of code coverage. 
% % \MYTODO{Dongge: Show how similar they are quantitatively e.g., differences are less then XXX percent. (1. Run a t-test on the final coverage scores, that p-value goes to here; 2. Compare their similarity across the whole 24h, that goes to Sec. 3)} 
% Specifically, the maximum difference between average branch coverage of all benchmarks is $1.15\%$, while the rest are all with $1\%$. 
% % This conclusion is also confirmed by a student t-test. 
% Our deep dive into the implementation of AFLNet and those algorithms revealed potential problems leading to those unexpected results. \MYTODO{Thuan: Problem 1: "compact" state machine, Problem 2: formulas to calculate the scores}.

To address the identified problems, in our second phase, we designed and implemented a novel state selection algorithm namely \AFLNetLegion based on Legion \cite{Legion}, which is a successful variant of the famous Monte Carlo tree search  \cite{MCTS}. We evaluated \AFLNetLegion in comparison with the existing \AFLNet algorithms. 
% \MYTODO{Thuan: At a high level, why MCTS was selected}. 
% The final results show that \MYTODO{Thuan: summarize the results}. Key insights extracted from the results \MYTODO{Thuan: fuzzing speed issue -- cite Marcel FSE'15 paper -- random fuzzing beats supposed-to-be effective symbolic execution due to its efficiency/speed, relationship between state selection, mutation algorithms, and state identification approaches}
The final results show that \AFLNetLegion clearly outperforms the baseline algorithms in selected case studies, demonstrating its principled design has advantage on programs that extend the motivating example.
However, the overall improvement is insignificant. These are unexpected yet interesting findings.

By analyzing the fuzzing artefacts, we identified two potential problems behind these unexpected results. The most critical problem could be the low fuzzing throughput (i.e., number of executions per second)) of \AFLNet. On average, it achieved only 20 execs/s, which is hundreds of times lower than the normal throughput achieved by \emph{stateless} fuzzers like AFL \cite{afl}. With such a low throughput, systematic algorithms like \AFLNetLegion might not be able to fully unlock its potential.

This paper makes the following contributions:
\begin{itemize}
  \item We conducted the first study, to the best of our knowledge, to evaluate an extensive set of state selection algorithms for stateful network protocol fuzzing. By analyzing the results, we identify problems and share insights that could open opportunities for future research in this topic.
  \item  We extended \Legion's algorithm to protocol state selection and made it available to support future research.
  % We extended \Legion's algorithm to protocol state selection and made it available to support future research.
\end{itemize}

\section{Background and Related Work}
\label{sec: aflnetlegion - background}

\subsection{Stateful Network Protocol Fuzzing}
% \MYTODO{Thuan: Complete the following tasks}
% \begin{itemize}
%   \item Quickly introduce fuzzing and advice reader to read TSE'19 paper for more information.
%   \item Introduce a general design for all types of stateful network protocol fuzzing, including black-box, white-box, and grey-box
%   \item Transition to AFLNet since it is the selected base fuzzer - more details for it is necessary
% \end{itemize}

\begin{figure}
  \centering\includegraphics[width=0.5\textwidth]{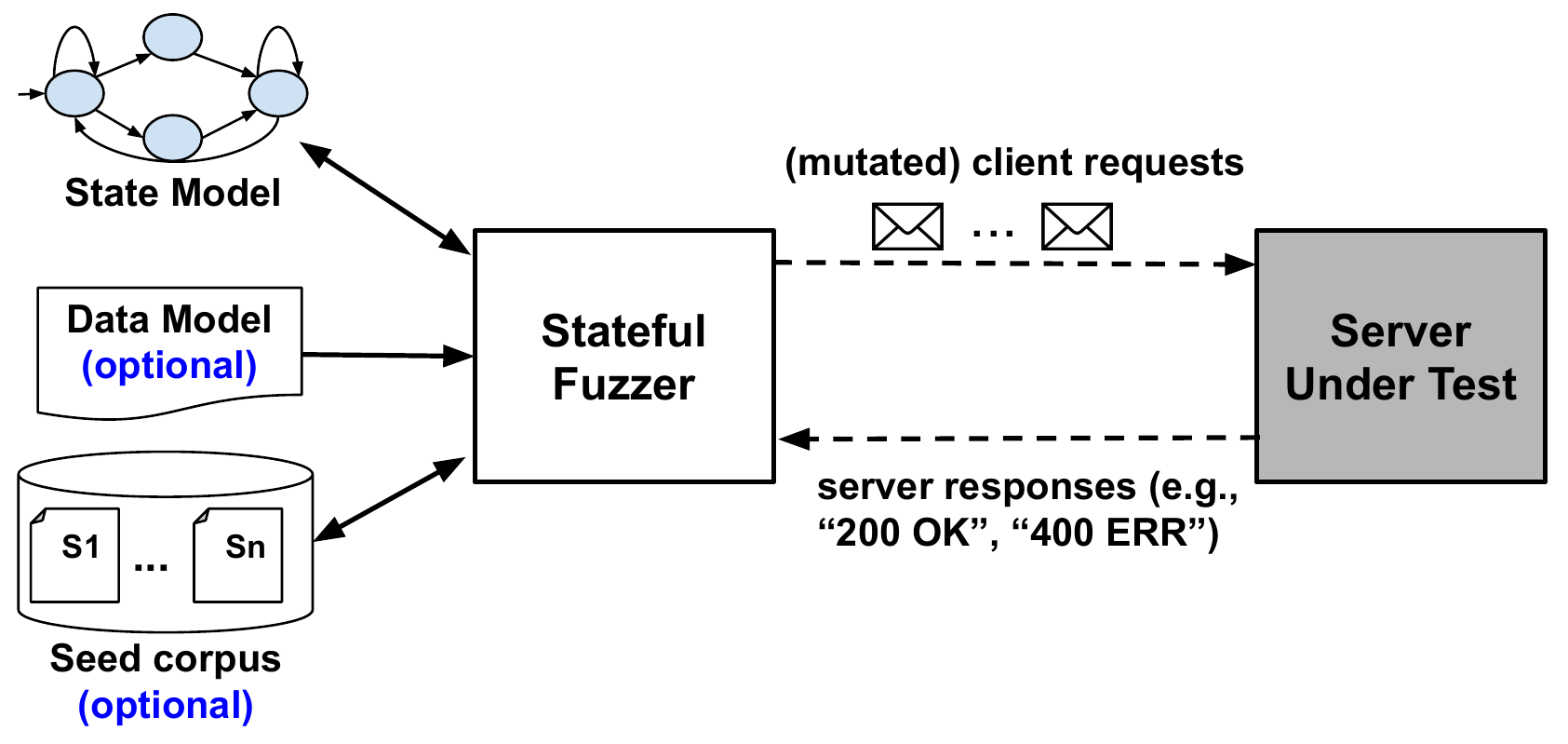}
  %\vspace{-5mm}
  \caption{Stateful Network Protocol Fuzzers}
  \label{fig:sfuzzers}
  %\vspace{-1em}
\end{figure}

Fuzzing is a process of repeatedly generating (random) inputs and feeding them to the system under test to cover more code and discover abnormal behaviours (e.g., crashes, memory access violations) \cite{ManesSurvey}. Since 1989, when the term fuzzing was coined by Prof. Barton Miller \cite{MillerFuzz}, it has received tremendous attention from both industry and academia, demonstrated by hundreds of published papers at top venues in Security and Software Engineering \cite{ManesSurvey}. The majority of these papers have focused on fuzzing stateless systems, whose behaviour depends only on the current input, such as media processing libraries and utility programs \cite{magma, fuzzbench}. The \emph{statefulness} property of network protocol servers pose a unique challenge for fuzzing.

To fuzz test stateful network servers, many techniques have been proposed including black-box, grey-box, and white-box approaches \cite{peach, boofuzz, aflnet, mace, prisma, pulsar, kif}, as depicted in Figure~\ref{fig:sfuzzers}. Unlike stateless fuzzing approaches, statefull fuzzers rely on a \emph{state model} that partitions the state space of the SUT to assist the test generation process. By leveraging the state model and monitoring the SUT's responses, the fuzzer knows what is the current SUT state and generates messages, which can be valid or invalid with respect to the message format expected by the SUT at that state. These messages can be generated in a fully random way~\cite{MillerFuzz}, from existing messages via mutations~\cite{aflnet}, or from a given data model/input grammar~\cite{peach, boofuzz}. Since not all states are equally important, the fuzzer also consults the state model to decide which states it should focus on. The decision is informed by \emph{state selection algorithms}.

\subsection{State Modelling Algorithms}
\label{sec: aflnetlegion - background - modelling}

Existing techniques tend to model state space using state machines. A state machine is a graph consisting of a finite number of states and transitions between states \cite{DiscreteMA}. The server response codes (e.g., "200 OK", "400 ERR") are commonly used to abstract the server states. Many works use a simple state machine,
where the destination of each transition is solely determined by the transition action and the origin state of the transition, regardless of other preceding states before the origin. For example, \textsc{KiF} \cite{kif} encodes a SIP system to a state machine; \textsc{snooze} and \textsc{mace} \cite{snooze,mace} uses it to record network operation information; and \textsc{tlshacker} \cite{protocolstatefuzzing,tlsattacker} analysed state machine models of TLS protocol implementations.
\textsc{PRISMA} \cite{prisma} defines the state machine model of the SUT in a probabilistic manner
with a hidden Markov model \cite{hiddenmm},
which further labels each transition with its execution probability. 
% \AFLNetLegion's tree model even more rigorously 
% determines the next server state with all preceding states.

To construct state models, many stateful fuzzing techniques rely on manually written models \cite{peach, boofuzz, sulley, snooze}, which requires the specification of the network protocols. Although this improves the probability of generating valid messages, the reliability of the model is subject to developers' understanding of the model and how well the specification describes actual implementations.

The state models can also be automatically constructed via either offline or online model learning. Offline learning is designed to infer a state model on a sufficient set of network traces and hence its accuracy relies heavily on that set. \textsc{PRISMA} \cite{prisma} proposed to cluster samples of network traffic to infer a state machine of the SUT, which also abstracts the message template (a.k.a message format) from the sample traffic. \textsc{PULSAR} \cite{pulsar} combines offline learning with fuzzing so that it can
generate messages from fuzzing templates with the state machine model and its embedded message rules.

% Online learning is proposed as an alternative to offline learning, it can learn and refine the server state model at runtime. A large message corpus can improve its performance,
% but is no longer indispensable to initiate fuzzing.
Online learning is proposed as an improvement on offline learning, it can learn and refine the server state model at runtime.
For example, 
Angluin’s L* algorithm \cite{angluinl*} is an online algorithm that has been applied in several research works \cite{protocolstatefuzzing,mace,prospex}
to gradually and dynamically infer a state machine model of the SUT through a number of messages.
As generative fuzzing techniques, their message generation requires the SUT-specific data model or message grammar. In particular, \cite{protocolstatefuzzing, mace} require a test harness to translate between abstract requests and actual packets; while \cite{prospex} proposes to learn it automatically.
% \cite{testdatagen} screens mutation
% One concern of generative fuzzing is that many messages generated may not be valid.
% An alternative to generate messages from data models is mutating existing traces extracted from real network traffic. \AFLNet \cite{aflnet} takes advantage of this so that it can produce valid new messages at a high probability and
% evolve from past interesting message sequences.
% Benefited from the same mutation-based strategy,
% \AFLNetLegion can start fuzzing from any server state preferred.

\subsection{State Selection Algorithms}
\label{sec: aflnetlegion - background - selection}

\textsc{Peach}, \textsc{BooFuzz}, \textsc{Snooze} \cite{peach, boofuzz, snooze} leave this burden to developers: developers have to embed their state selection strategies into the manually written state models. \textsc{PRISMA} \cite{prisma} corresponds each state to a template and randomly chooses one according to the transition probability of its Markov model. \textsc{PULSAR} \cite{pulsar} has multiple templates in one state and prefers templates with more fuzzable fields. \textsc{MACE} \cite{mace} uses a priority queue to favour states traversed by past executions that visited a large number of unexplored basic blocks.

\subsection{AFLNet: Stateful Grey-box Fuzzing}
\label{sec: aflnetlegion - background - aflnet}

\AFLNet is a greybox fuzzer for stateful protocol implementations. Unlike existing protocol fuzzers, it takes a mutational approach and uses state-feedback, in addition to code-coverage feedback, to guide the fuzzing process. AFLNet is seeded with a corpus of recorded message exchanges between the server and an actual client. No protocol specification or message grammars are required. It acts as a client and replays variations of the original sequence of messages sent to the server and retains those variations that were effective at increasing the coverage of the code or state space. To identify the server states that are exercised by a message sequence, AFLNet uses the server’s response codes. 

% From this feedback, AFLNet identifies progressive regions in the state space, and systematically steers towards such regions.

Regarding state selection policies, in its default state selection algorithm called FAVOR, \AFLNet \cite{aflnet} uses an intuitive formula to consider states that are less often targeted, trying to balance exploiting the benefit of states that appear interesting with exploring the potential of the ones that seem to be less progressive. It also supports two more state selection algorithms called RANDOM and ROUND-ROBIN. While the former algorithm randomly selects states, the latter stores states in a kind-of circular queue and selects them in turns.

% Since \AFLNet is a mutational fuzzing approach, it leverages existing seeds that are sequences of messages in stateful server fuzzing. 
Once a state is selected, \AFLNet chooses a seed input (i.e. a sequence of messages) that can reach that state. The seed selection algorithms are also named FAVOR, RANDOM, and ROUND-ROBIN. Like AFL \cite{afl}, the FAVOR seed selection algorithm prioritises seed inputs that are small, cover more code, and take less time to execute. Meanwhile, the RANDOM and ROUND-ROBIN algorithms follow the same logic of RANDOM and ROUND-ROBIN state selection algorithms as described above.

%subsection{Legion}
\subsection{Monte Carlo Tree Search and Legion}
\label{sec: aflnetlegion - background - monte carlo tree search and legion}
% \MYTODO{Dongge? Talk about background for MCTS first and the high-level idea of Legion as a successful application of MCTS}

The \textit{Monte Carlo tree search} (MCTS) algorithm \cite{MCTS} has proven its efficiency in exploring large search space,
such as the state space of complex games like Go \cite{silver2016alphago}.
It iteratively refines a tree model of the search space via four  steps: \emph{selection}, \emph{expansion}, \emph{simulation}, and \emph{back-propagation}.
The tree can facilitate a \textit{best-first} exploration policy,
as four steps can estimate the productivity of various exploration directions and dynamically adjust the estimation with the statistics of simulation results.

% while the four steps progressively refine the accuracy of tree's estimation. 
% evaluating by dynamically updating the value of each node tree
% adjust the exploration policy  based simulation results,
% It adjusts its exploration direction by adapting to the search space via iterations of four steps: namely .

% In each iteration,
% it starts with selection, where it descends down the tree from the root node and repeatedly selects the node with the highest value, until it reaches an expandable node.
% The value is determined by an \textit{Upper Confidence Tree} (UCT) function, based on multi-armed bandit learning, that 
% % optimistically evaluates each tree node, to
% not only prefers selecting nodes with high past rewards, but also nodes that appear to be under-explored.
% The expandable node is a non-terminal state that has at least one un-visited child.
% The second step, namely expansion, 
% takes a further action (e.g., the next move in Go) from the expandable node and adds the corresponding child to the node.
% From the new child, it start the third step: Simulation.
% This step randomly explores the rest of the search space from the state of the leaf node, until it reaches a terminal state (e.g., the end of the game Go).
% The outcome of simulation is considered as the reward to the latest selection process.
% The last step of each iteration is propagation, where it updates the statistics of all selected nodes with the new reward.

\Legion \cite{Legion} proposed a principled Monte Carlo tree search based algorithm
for coverage-guided ordinary software testing.
Its variation of MCTS attempts to learn the most promising program states to investigate at each search iteration based on past observation on code coverage.
% Like the original MCTS, this strategy exhibits satisfying performance. For example,
% it outperformed \texttt{KLEE} in 7 out of 11 benchmark suites from \texttt{TestComp}. 
\Legion considers the symbolic tree of the program as the search space, and explores it with four steps that are marginally varied from the original MCTS to adapt to concolic execution.

Its selection step descends down the tree until a \textit{simulation child} is selected.
A simulation child is added to each node of the original MCTS search tree to allow simulation
from their parent's program state, even if they are intermediate nodes.

Simulation is the second step in \Legion.
It first generates inputs that preserve the path selected 
by solving the path constraint of the simulation child's parent.
Simulating from intermediate nodes is designed to reduce the computation cost of constraint solving.
For example,
we do not have to solve the complex constraints of deep program states,
if solving their ancestors' simpler constraints turns out to be as productive.
\Legion will then execute the program under test with the inputs generated.
Departing from the original MCTS,
\Legion records the whole execution path to determine the outcome (reward) of the current selection in the next step.

The third step is expansion,
which checks if each new execution path exists in the current tree and adds the whole path to the tree.
It records a reward to the current selection if the path is new.
Each iteration ends with propagation.
It increases the selection count of all nodes in the selection path by one,
and adds the rewards to the discovery count of all nodes in the execution path.
Note that some execution paths may fail to preserve the selection path,
due to glitches in constraint solving in practice. 

The design of \Legion also benefits light-weight (i.e., without symbolic execution) stateful protocol fuzzing techniques like \AFLNet.
For example, simulation from intermediate tree nodes is necessary,
because when symbolic execution is not available,
we cannot confirm if all child protocol states of a node are covered.
\AFLNetLegion takes advantage of this design with its own modifications to adapt to protocol modelling and fuzzing,
we explain the detail in \cref{sec: aflnetlegion - approach}.

\section{Preliminary Study}
\label{sec: aflnet - preliminary}
% \MYTODO{Dongge and Thuan: Complete the following task}
% \begin{itemize}
%     \item Short explanation of the setup
%     \item [x] Results & analysis (Dongge)
%     \item Transition to the next section
% \end{itemize}
In this first phase of our investigation, we evaluated the three existing state selection algorithms namely RANDOM, ROUND-ROBIN, and FAVOR of \AFLNet. We discussed the details of these algorithms in \cref{sec: aflnetlegion - background - aflnet}.  

\subsection{Experimental Setup}

\iparagraph{Benchmarking programs.} We conducted this preliminary study on six subjects (\cref{tab:subjects}) from ProFuzzBench \cite{profuzzbench}, the largest public benchmark for network protocol fuzzers.
We intentionally selected these \emph{stateful} protocols for evaluation, because they are consistent with this paper's interest, namely the impact of selection algorithms on fuzzing stateful protocol implementations. Moreover, their setups produce stable results.
% For the same reason, we did not include other protocols supported by ProFuzzBench.

%\iparagraph{Benchmarking programs.}

\begin{table}[h!]
\vspace{1em}
\centering
\small
\begin{tabular}{|l|l|l|}
  \hline
  Subject                & Protocol     & Description \\ \hline\hline
  LightFTP  & FTP   & File Transfer Protocol \\ \hline
  ProFTPD  & FTP   & File Transfer Protocol \\ \hline 
  Exim  & SMTP   & Simple Mail Transfer Protocol \\ \hline
  OpenSSH  & SSH   & Secure Remote Login and File Transfer \\ \hline
  OpenSSL  & TLS   &   Secure socket connection \\
  \hline
  Live555  & RTSP   &  Real-time media streaming \\ \hline
\end{tabular}
\vspace{0.5em}
\caption{\label{tab:subjects}Subject programs}
\end{table}
\vspace{0.2em}

\iparagraph{Seeds, Timeout and Repetition.}
We used the seed corpora provided by the ProFuzzBench benchmark. Regarding timeout, we followed the suggestions from \cite{klees2018evaluating} and ran our experiments for 24 hours. We also repeated the experiments 5 times to mitigate the influence of randomness.
% {\color{red} Dongge, I think it is fine to say that for all subjects we ran 5 trials, and we ran 40 trials for the case studies. Otherwise, it is difficult to explain}.

\iparagraph{Performance Measurement.}
Since \AFLNet is a coverage-based approach,
we focus on reporting its achieved code coverage. There is a (arguably) common understanding that covering more code leads to higher chance of discovering software faults. We used ProFuzzBench's scripts to fully automate the execution of the target servers inside Docker containers, and generate code coverage reports.

\iparagraph{Platform.}
All experiments were conducted on the same host machine, running Ubuntu 18.04 64-bit with a 2.50 GHz Intel Xeon Platinum 8180M CPU and 128 GB of RAM.

\subsection{Preliminary Results and Analysis}
% {\color{gray}
% Compare the performance of 3 \AFLNet algorithms;
% }
\begin{figure*}[ht!]
    \centering
    \begin{subfigure}[b]{0.32\textwidth}
        \centering
        \includegraphics[width=\textwidth]{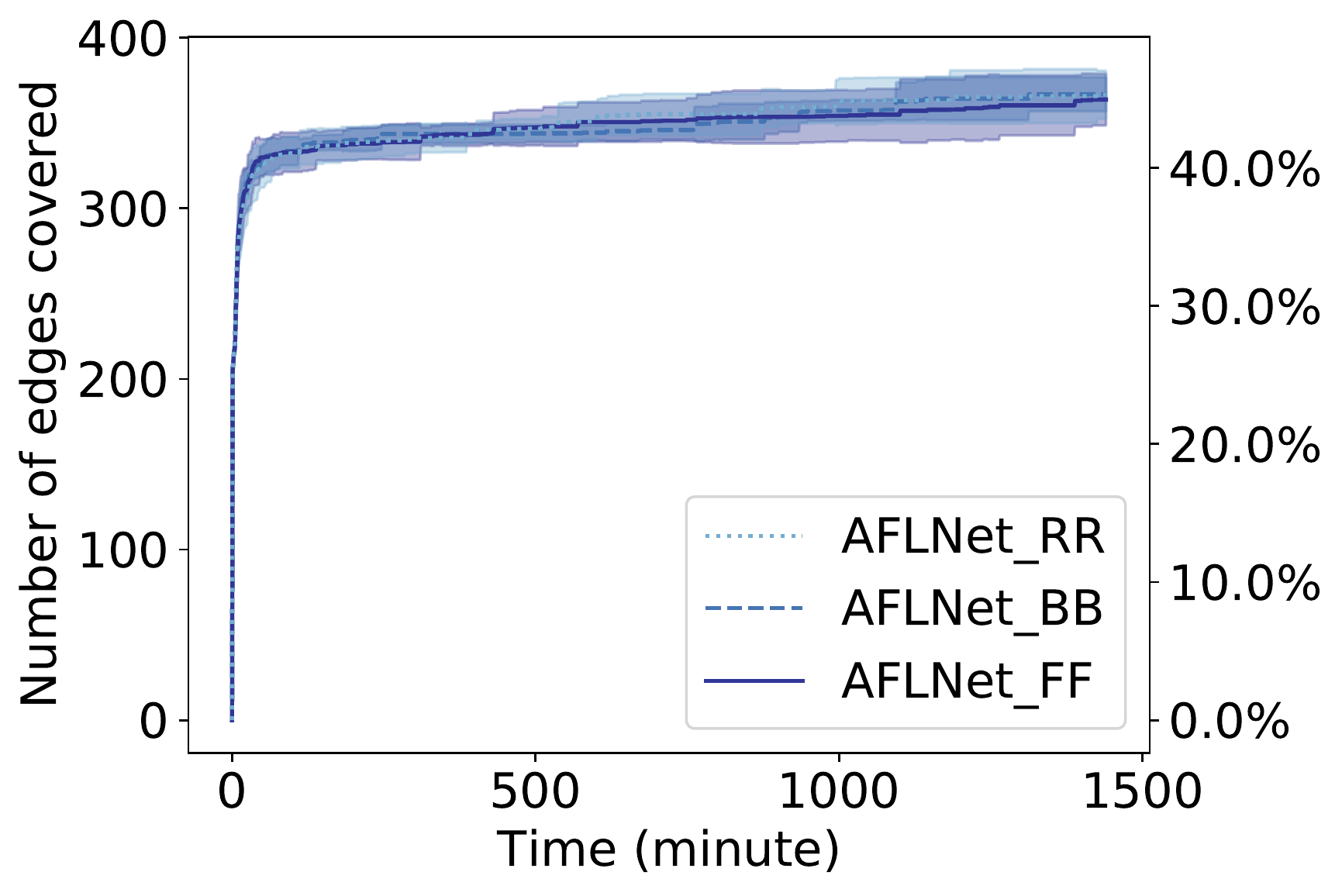}
        \caption{\texttt{LightFTP} branch coverage}
        \label{fig: aflnet overall lightftp branch coverage}
    \end{subfigure}
    \hfill
    \begin{subfigure}[b]{0.32\textwidth}
        \centering
        \includegraphics[width=\textwidth]{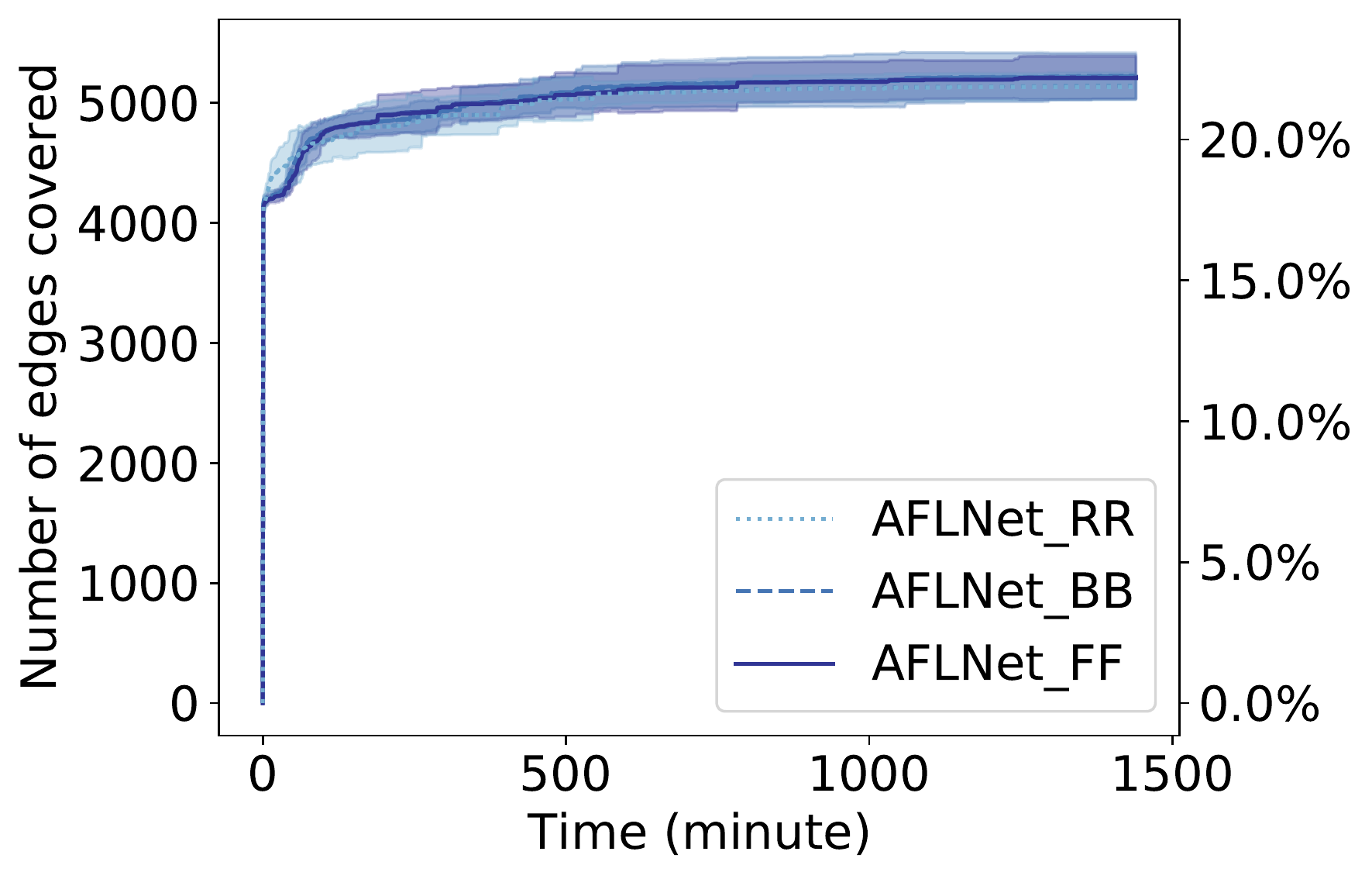}
        \caption{\texttt{ProFTPD} branch coverage}
        \label{fig: aflnet overall proftpd branch coverage}
    \end{subfigure}
    \hfill
    \begin{subfigure}[b]{0.32\textwidth}
        \centering
        \includegraphics[width=\textwidth]{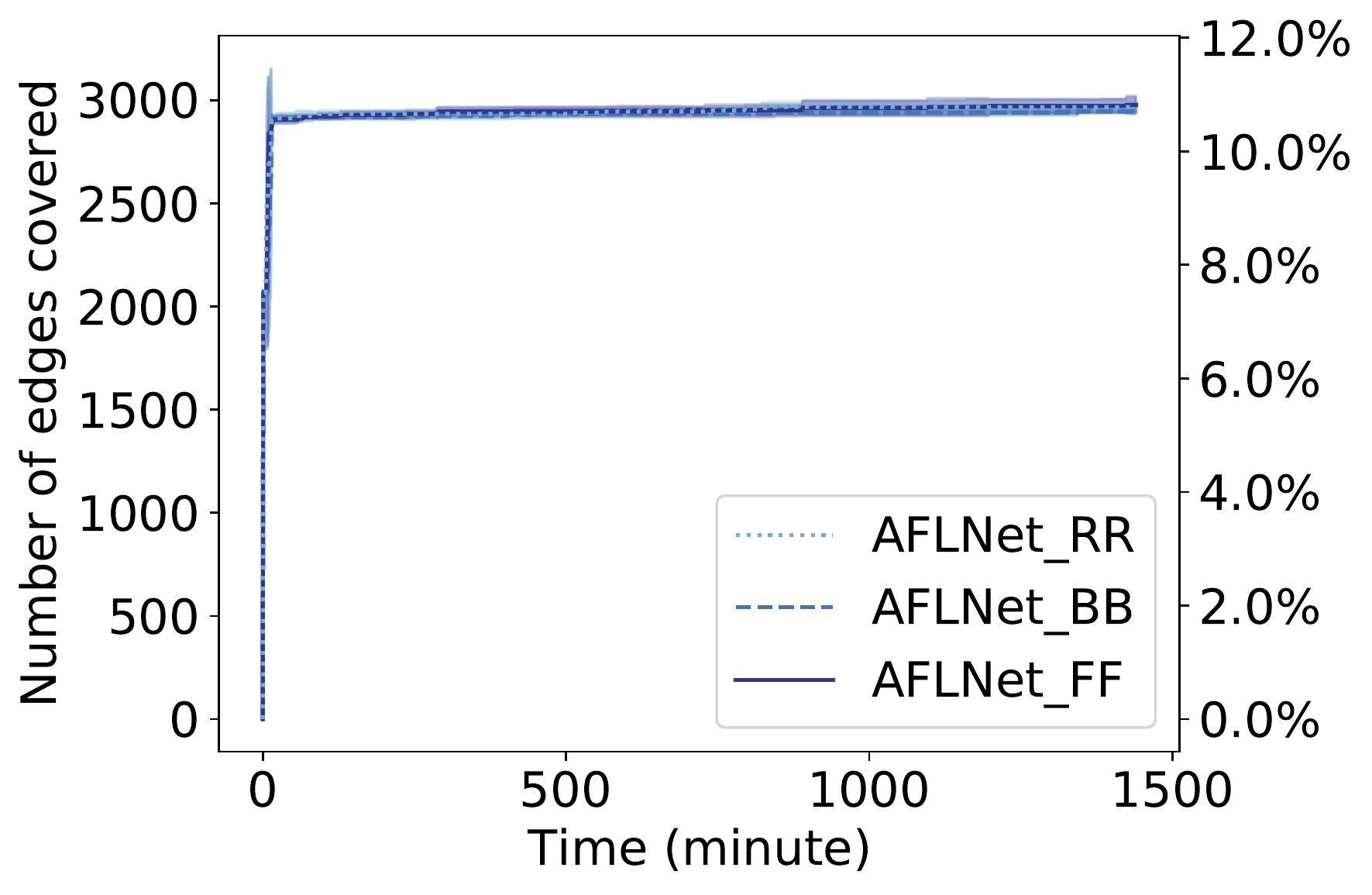}
        \caption{\texttt{Exim} branch coverage}
        \label{fig: aflnet overall exim branch coverage}
    \end{subfigure}
    % \vskip\baselineskip
    % \begin{subfigure}[b]{0.32\textwidth}
    %     \centering
    %     \includegraphics[width=\textwidth]{Figures/lightftp/l_aflnet.pdf}
    %     \caption{\texttt{LightFTP} line coverage}
    %     \label{fig: overall lightftp line coverage}
    % \end{subfigure}
    % \hfill
    % \begin{subfigure}[b]{0.32\textwidth}
    %     \centering
    %     \includegraphics[width=\textwidth]{Figures/proftpd/l_aflnet.pdf}
    %     \caption{\texttt{ProFTPD} line coverage}
    %     \label{fig: overall proftpd line coverage}
    % \end{subfigure}
    % \hfill
    % \begin{subfigure}[b]{0.32\textwidth}
    %     \centering
    %     \includegraphics[width=\textwidth]{Figures/exim/l_aflnet.pdf}
    %     \caption{\texttt{Exim} line coverage}
    %     \label{fig: overall exim line coverage}
    % \end{subfigure}
\end{figure*}
\begin{figure*}[ht!]\ContinuedFloat
    \begin{subfigure}[b]{0.32\textwidth}
        \centering
        \includegraphics[width=\textwidth]{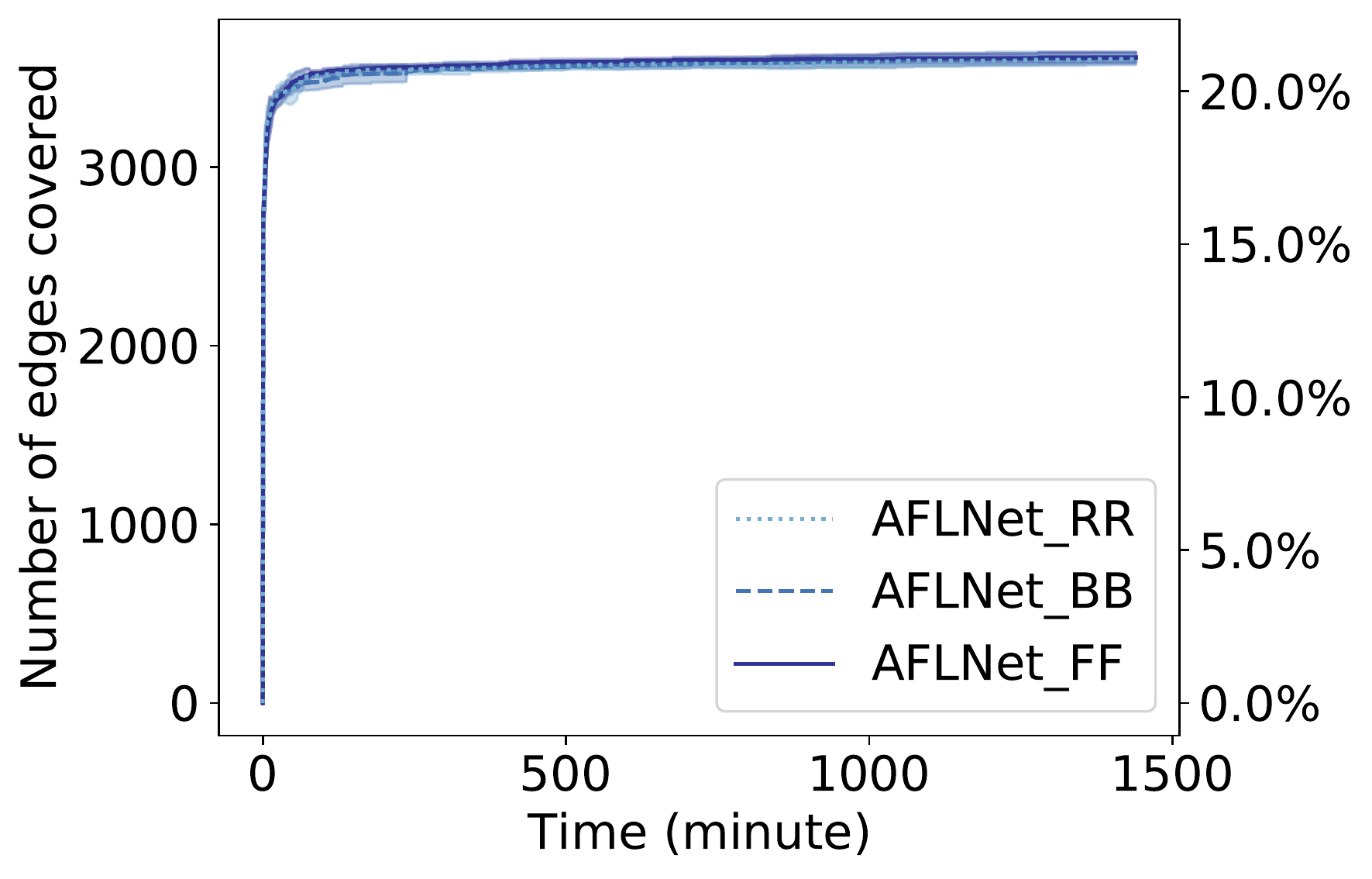}
        \caption{\texttt{OpenSSH} branch coverage}
        \label{fig: aflnet overall openssh branch coverage}
    \end{subfigure}
    \hfill
    \begin{subfigure}[b]{0.32\textwidth}
        \centering
        \includegraphics[width=\textwidth]{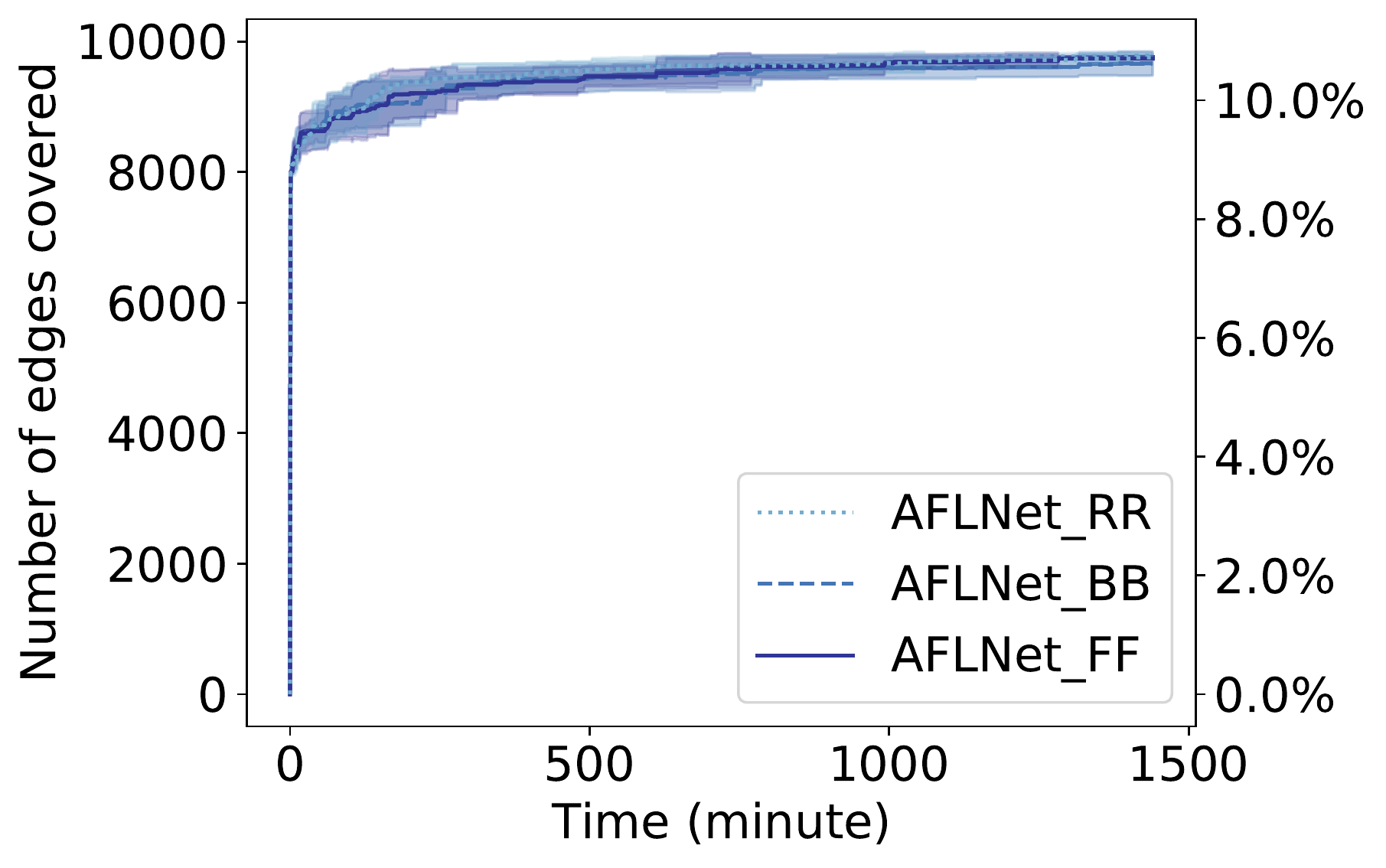}
        \caption{\texttt{OpenSSL} branch coverage}
        \label{fig: aflnet overall openssl branch coverage}
    \end{subfigure}
    \hfill
    \begin{subfigure}[b]{0.32\textwidth}
        \centering
        \includegraphics[width=\textwidth]{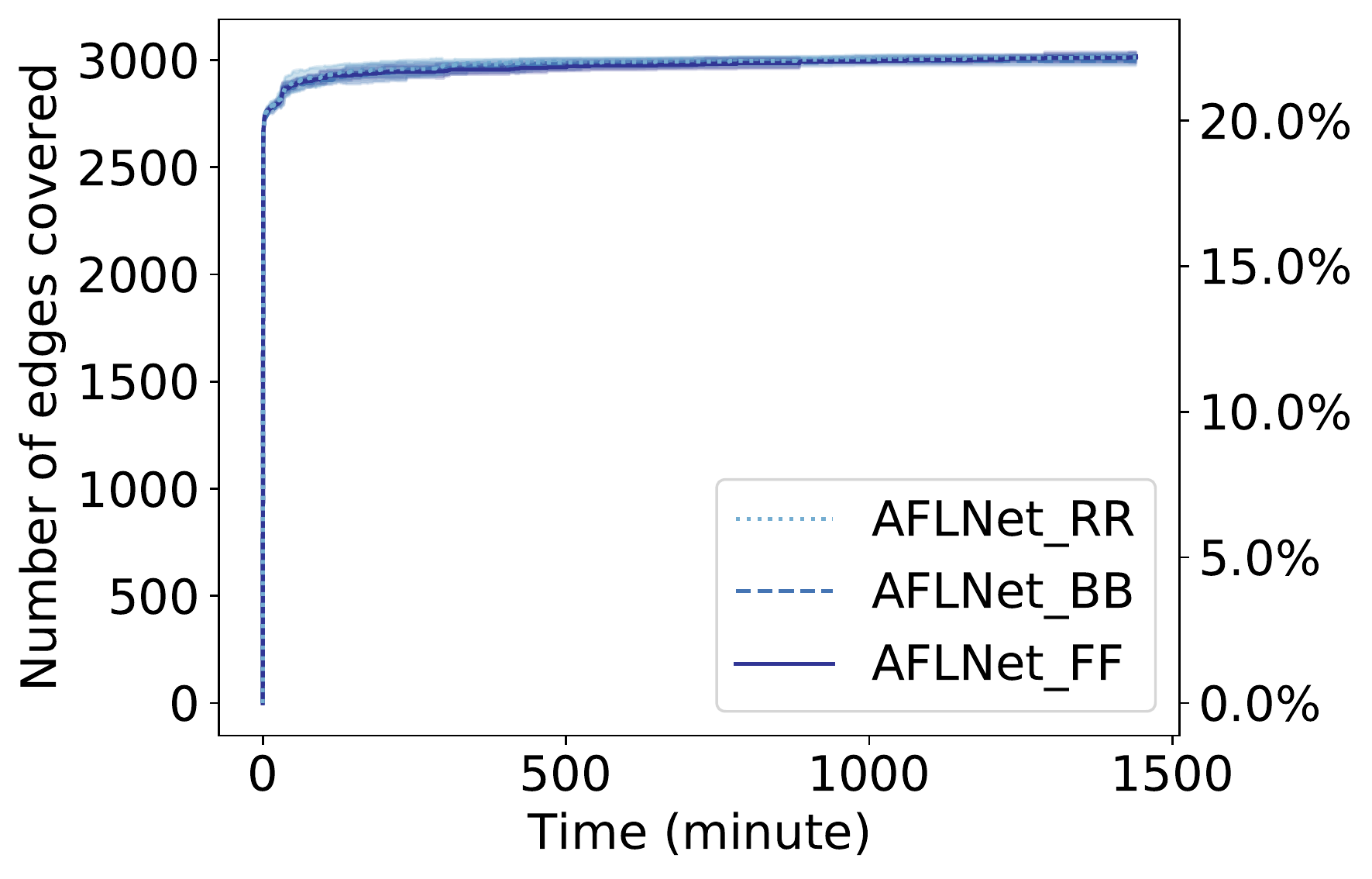}
        \caption{\texttt{Live555} branch coverage}
        \label{fig: aflnet overall live555 branch coverage}
    \end{subfigure}
    % \vskip\baselineskip
    % \begin{subfigure}[b]{0.32\textwidth}
    %     \centering
    %     \includegraphics[width=\textwidth]{Figures/openssh/l_aflnet.pdf}
    %     \caption{\texttt{OpenSSH} line coverage}
    %     \label{fig: overall openssh line coverage}
    % \end{subfigure}
    % \hfill
    % \begin{subfigure}[b]{0.32\textwidth}
    %     \centering
    %     \includegraphics[width=\textwidth]{Figures/openssl/l_aflnet.pdf}
    %     \caption{\texttt{OpenSSL} line coverage}
    %     \label{fig: overall openssl line coverage}
    % \end{subfigure}
    % \hfill
    % \begin{subfigure}[b]{0.32\textwidth}
    %     \centering
    %     \includegraphics[width=\textwidth]{Figures/live555/l_aflnet.pdf}
    %     \caption{\texttt{Live555} line coverage}
    %     \label{fig: overall live555 line coverage}
    % \end{subfigure}
    \caption{\AFLNet Overall results}
    \label{fig: aflnet overall results}
\end{figure*}

Our experiments show that
distinct state selection algorithms exhibit minimum impact on the coverage performance.
\cref{fig: aflnet overall results} plots 
three algorithms' average absolute and percentage branch coverage performance (respectively on the left and the right y-axes) on all benchmark subjects.
% Recall from \cref{sec: aflnetlegion - background}
% that the three state selection algorithms are respectively
% random (\aflnetrr, in light blue dotted lines),
% round robin (\aflnetbb, in blue dashed lines),
% and favour (\aflnetff, in dark blue solid lines).
The RANDOM, ROUND-ROBIN, and FAVOUR algorithms are respectively denoted by
\aflnetrr, \aflnetbb, and \aflnetff,
drawn in
light blue dotted lines, blue dashed lines, and dark blue solid lines.
Each mean curve is surrounded by its $95\%$ confidence interval.

In general,
all algorithms tend to exhibit similar coverage performance
(indicated by their overlapping curves)
and stability
(shown by the similar widths of their confidence intervals)
on each benchmark.
For each subject,
the differences between all algorithms' average final branch and live coverage values are within $1.2\%$.
% The only exception is that
% FAVOUR outperformed RANDOM baseline by $1.15\%$ in branch coverage \texttt{ProFTPD}
A 2-sample Student's t-test \cite{student1908probable} further confirmed that such differences between their performance are statistically insignificant.
% which shows the p-values of almost all final mean coverage results are greater than $5\%$,
% except the branch coverage on \texttt{ProFTPD} between FAVOUR and RANDOM.

In addition to the close final coverage results,
all algorithms also take very similar coverage trajectories on each subject.
At the very beginning,
all algorithms can quickly and stably cover similar amount of branches and lines
with very tight confidence intervals.
% illustrated by the almost vertical line in the first few minutes.\MYTODO{Thuan: discuss with Dongge}
Then diverged performance starts to emerge
between different algorithms (i.e., larger distances between mean curves)
and between trials of the same algorithm (i.e. wider confidence intervals).
Take benchmark subject \texttt{ProFTP} as an example,
the RANDOM baseline outperformed the carefully designed FAVOUR for a short period
immediately after the initial stage.
% in both \cref{fig: overall proftpd branch coverage} and \cref{fig: overall  proftpd line coverage}.
Similarly,
in \texttt{Exim}, the confidence interval of RANDOM expanded for a very short period,
% before quickly decreasing,
resulting in a spike-like-region at the top-left corner of
\cref{fig: aflnet overall exim branch coverage}.
% and \cref{fig: overall exim line coverage}.
% Note that this does not mean some RANDOM trials covered more than $3000$ branches.
% Instead, it is because some random algorithms covered much less branches
% than average, causing a large standard deviation and confidence interval at those timestamps.
Note that it is a fluctuation of the confidence interval (rather than mean) of RANDOM, caused by some trials covering much fewer branches than average at those timestamps.
Nonetheless,
the differences between curves are almost always within the confidence interval,
indicating the divergence between algorithms is insignificant
compared with the differences between trials of the same algorithm.

It is worth noting that, at the end of most experiments, three algorithms' curves visually overlap on each other with a tight confidence interval,
implying their stably close performances.
Two exceptions are the \texttt{FTP} benchmarks,
where in \texttt{LightFTP}
(\cref{fig: aflnet overall lightftp branch coverage})
% ,\cref{fig: overall lightftp line coverage})
RANDOM performed closely to ROUND-ROBIN and outperformed FAVOUR,
and in \texttt{ProFTPD}
(\cref{fig: aflnet overall proftpd branch coverage})
% \aflnetbb and \aflnetff slightly outperformed \aflnetrr
the performances of all were not as stable as the other benchmarks
across the whole 24 hour experiment.

%MYTODO{A transition to the next section}

The indistinguishable coverage performance between FAVOUR and the other two distinct alternatives raises the following question:\\
\emph{Is the unexpected result caused by some weaknesses of FAVOUR?}\\
In \cref{sec: aflnetlegion - approach}, we identify two main weaknesses of \AFLNet and then propose corresponding fixes with a principled selection algorithm. 

% \MYTODO{Move to Sec. V:
% \cref{sec: aflnetlegion - evaluation} will further prove that even  the principled selection algorithm cannot largely improve the performance either.
% }

\section{MCTS-based State Selection Algorithm}
\label{sec: aflnetlegion - approach}

%\MYTODO{Connect to the previous section.}
% \MYTODO{Thuan? Complete the following task}
% \begin{itemize}
%     \item Motivating example that shows the problems of AFLNet and the advantages of MCTS/Legion
%     \item Move the algorithm section here & Update the text accordingly
% \end{itemize}

To prove the unexpected results are irrelevant to the weaknesses in \AFLNet,
this section proposes a principled selection algorithm to address two main weaknesses of \AFLNet that may limit its performance.
We term the new algorithm \AFLNetLegion, as it extends \Legion's algorithm to stateful protocol fuzzing on the basis of \AFLNet.

We identify the following two main weaknesses in \AFLNet's state selection algorithm.
First, its coarse-grained state machine model identifies each state
only by individual response codes without distinguishing their prefixes,
leading to unreliable state evaluation and selection.
% complicates request sequence selection within each state and 
Second, its intuitive state evaluation formula 
lacks a principled way to balance the exploration-exploitation trade-off,
which may inaccurate estimate the productivity of server states.

\cref{subsec: motivating example} exemplifies how the compact state model of \AFLNet may limit the performance of its state selection algorithm. \cref{sec: algorithm} explains why it can be improved with a more informative tree model inspired by \Legion. \cref{subsec: tree nodes} and \cref{subsec: tree construction} describe the construction detail of the tree model. \cref{subsec: selection policy} then introduces the principled UCT function used by \Legion as an improvement to the intuitive state evaluation formula of \AFLNet.

% \MYTODO{Thuan: the motivating example talks a lot about how different models may affect the accuracy of state selection, but very little about the problem with the current \AFLNet's state evaluation algorithm.
% Maybe we can merge them?
% In the past we wrote them in parallel,
% but that makes it hard to tell the story without breaking the logic flow.
% My current strategy is to explain them in a sequence. 
% We start by only talking about the state model 
% \emph{as a way to assist state selection}
% in
% \cref{subsec: motivating example},
% \cref{sec: algorithm},
% \cref{subsec: tree nodes} and \cref{subsec: tree construction}.
% Then we very briefly mention the UCT function in the selection algorithm is directly borrowed from \Legion in \cref{subsec: selection policy}.
% }

% \MYTODO{Maybe give a high level description of the two weaknesses here?}

\subsection{Motivating Example} \label{subsec: motivating example}

% Second, its state selection algorithm lacks
% a systematic and principled way to balance 
% in the large and complex server state space,
% under the uncertainty of the external server implementation.

\begin{figure*}[ht]
    
    \begin{subfigure}[b]{0.49\textwidth}
    \centering
\begin{lstlisting}[style=aflnet]
USER foo
PASS bar
NLST *  #Globbing matches nothing
MKD  temp
STOR sample.text
\end{lstlisting}
    \caption{Request sequence 1: \texttt{Globbing} before creating directory and file}
    \label{lst: request sequence 1}
    \end{subfigure}
    \hfill
    \begin{subfigure}[b]{0.49\textwidth}
\begin{lstlisting}[style=aflnet, emph={NLST}]
USER foo
PASS bar
MKD  temp
STOR sample.text
NLST *  #Globbing matches new directory and file
\end{lstlisting}
    \caption{Request sequence 2: \texttt{Globbing} after creating directory and file}
    \label{lst: request sequence 2}
    \end{subfigure}
    \caption{Two sample request sequences involving globbing with \texttt{NLST}}
    \label{fig: RequestSeqs}
\end{figure*}

We will start with an example that spotlights the weaknesses of \AFLNet and the corresponding improvements allowed by \AFLNetLegion.
\cref{fig: RequestSeqs} is a case study illustrating how 
the effect of its state selection algorithm
can be diminished by
a coarse-grained
% input space partition from the 
state machine model.
% how a principle approach from \Legion mitigates these two weaknesses.
It uses the Name List (\NLST) FTP command to explain the importance of selecting the \emph{interesting} state in stateful protocol fuzzing.
The command will instruct the server to list the file names that match with its parameter.
Here we use an asterisk (\texttt{*}) as its parameter to test the \glob library.
The library is in charge of matching files and directories with wildcard characters (e.g. \texttt{*}) or other patterns.
In this example, two sequences send the same requests in different orders:
Sequence 1 (\cref{lst: request sequence 1}) on the left globs with a \NLST request without the existence of any file or directory;
Sequence 2 (\cref{lst: request sequence 2}) on the right creates a directory and a file before sending the same \NLST command.
As a result,
the \NLST command in sequence 2 is more interesting than sequence 1 to fuzz, because its mutants are more probable to cover more code of the \glob library
when the server attempts to match the parameter (e.g. *).
A rigorous model of the server should distinguish these two sequences
so that the selection algorithm can later learn that the state of sequence 2 is more productive in fuzzing than 1.
% This makes it a typical example where
% the performance of state selection algorithms can be limited
% by \AFLNet's compact coarse-grained state machine model.

\begin{figure*}[ht!]
  \centering
  \begin{subfigure}[b]{0.49\textwidth}
      \centering
      \begin{adjustbox}{width=\textwidth}
    \begin {tikzpicture}[
        auto, on grid, semithick,
        state/.style = {
            circle, top color=white, bottom color=white,
            node distance = 2cm and 2cm,
            draw, black, text=black,
            minimum size = 1cm,
            },
        el/.style = {inner sep=2pt, align=left, sloped},
        every label/.append style = {font=\tiny}
    ]

        \node[state] (0)   []               {0};
        \node[state] (331) [right = of 0]  {331};
        \node[state] (230) [right = of 331] {230};
        \node[state] (250) [right = of 230] {250};
        \node[state] (257) [right = of 250] {257};

        \path[->] (0)   edge [] node[el, above] {\texttt{USER}} (331);
        \path[->] (331) edge [] node[el, above] {\texttt{PASS}} (230);
        \path[->] (230) edge [] node[el, above] {\texttt{NLST}} (250);
        \path[->] (230) edge [bend right=50] node[el, above] {\texttt{MKD}} (257);
        \path[->] (250) edge [in=120,out=60,loop] node[el, above] {\texttt{\textbf{NLST}}} (250);
        \path[->] (250) edge [bend right=20] node[el, below] {\texttt{MKD}} (257);
        \path[->] (257) edge [bend right=20] node[el, above] {\texttt{STOR}} (250);

        % \draw[->] (0)   -- (331) node [midway, label=above:{\texttt{USER}}] {};
        % \draw[->] (331) -- (230) node [midway, label=above:{\texttt{PASS}}] {};
        % \draw[->] (230) -- (250) node [midway, label=above:{\texttt{NLST}}] {};
        % \draw[->] (230) -- (257) node [midway, label=above:{\texttt{MKD}}] {};
        % \draw[->] (250) -- (250) node [midway, label=above:{\texttt{NLST}}] {};
        % \draw[->] (250) -- (257) node [midway, label=above:{\texttt{MKD}}] {};
        % \draw[->] (257) -- (250) node [midway, label=above:{\texttt{STOR}}] {};

    \end{tikzpicture}
\end{adjustbox}
      \vspace*{-.3cm}
      \caption{\AFLNet's state machine model}
      \label{fig: AFLNet state machine}
  \end{subfigure}
  \hfill
  \begin{subfigure}[b]{0.49\textwidth}
      \centering
      \begin{adjustbox}{width=\textwidth}
    \begin {tikzpicture}[
        auto, on grid, semithick,
        state/.style = {
            circle, top color=white, bottom color=white,
            node distance = 2cm and 2cm,
            draw, black, text=black,
            minimum size = 1cm,
            },
        el/.style = {inner sep=2pt, align=left, sloped},
        every label/.append style = {font=\tiny}
    ]

        \node[state] (0)   []               {0};
        \node[state] (331) [right = of 0]  {331};
        \node[state] (230) [right = of 331] {230};
        \node[state] (250) [right = of 230] {250};
        \node[state] (257) [right = of 250] {257};
        \node[state] (250') [right = of 257] {250'};
        \node[state] (257') [below = of 250] {257'};
        \node[state] (250'') [right = of 257'] {250''};
        \node[state] (250''') [right = of 250''] {250'''};

        \path[->] (0)   edge [] node[el, above] {\texttt{USER}} (331);
        \path[->] (331) edge [] node[el, above] {\texttt{PASS}} (230);
        \path[->] (230) edge [] node[el, above] {\texttt{NLST}} (250);
        \path[->] (250) edge [] node[el, above] {\texttt{MKD}} (257);
        \path[->] (257) edge [] node[el, above] {\texttt{STOR}} (250');
        \path[->] (230) edge [] node[el, above] {\texttt{MKD}} (257');
        \path[->] (257') edge [] node[el, above] {\texttt{STOR}} (250'');
        \path[->] (250'') edge [] node[el, above] {\texttt{\textbf{NLST}}} (250''');

        % \draw[->] (0)   -- (331) node [midway, label=above:{\texttt{USER}}] {};
        % \draw[->] (331) -- (230) node [midway, label=above:{\texttt{PASS}}] {};
        % \draw[->] (230) -- (250) node [midway, label=above:{\texttt{NLST}}] {};
        % \draw[->] (230) -- (257) node [midway, label=above:{\texttt{MKD}}] {};
        % \draw[->] (250) -- (250) node [midway, label=above:{\texttt{NLST}}] {};
        % \draw[->] (250) -- (257) node [midway, label=above:{\texttt{MKD}}] {};
        % \draw[->] (257) -- (250) node [midway, label=above:{\texttt{STOR}}] {};

    \end{tikzpicture}
\end{adjustbox}
      \vspace*{-.3cm}
      \caption{\AFLNetLegion's tree model}
      \label{fig: AFLNetLegion tree}
  \end{subfigure}
  \caption{Two models of the request sequences in \cref{fig: RequestSeqs}\\
          the bold font \textbf{\texttt{NLST}} is the interesting one from \AFLNetLegion}
\end{figure*}
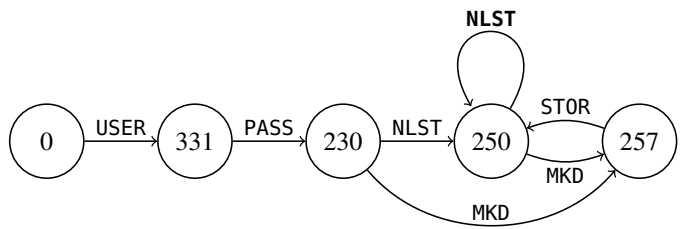
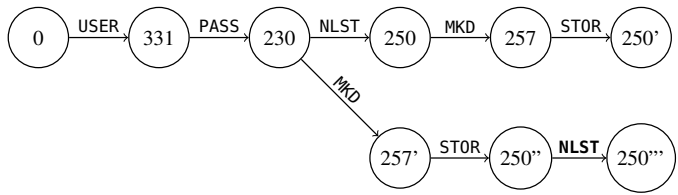

\cref{fig: AFLNet state machine} illustrates the state machine of
\AFLNet after receiving the server's responses to
two sequences of requests.
The initial server state is represented by a dummy code 0,
and each following state is identified with the response code of the latest command.
% The blue and red line respectively represent the response sequence from sequence 1 and 2.
Note that the \texttt{FTP} server returns the same code (\texttt{250}) to successful 
\NLST and \texttt{STOR} commands,
hence \AFLNet's state machine is unable to distinguish them,
and will think both request sequences produce the same state (\texttt{250}),
where it can send the interesting \NLST from sequence 2.
Similarly, with more request sequences generated later,
it may further mislead the selection algorithm to consider more noise request sequences
as the prefixes of the interesting \NLST,
such as \texttt{USER -> PASS -> NLST}.
% In addition,
% it may also complicates request sequence selection after a state is selected,
% e.g. when selecting a \NLST from \texttt{USER -> PASS -> NLST -> NLST -> MKD -> STOR -> NLST}.
% 
% \AFLNet's state machine models limit the performance of selection algorithms,
% because it partitions the input space at a coarse-gain level.
% For example, 
% As a result,
% this model groups both request sequences in state \texttt{250}.
As a result,
the boring noise request sequence 1 lowers the estimated value of the state (\texttt{250})
and consequently reduces the chance of selecting the interesting state from sequence 2.
% In addition, in the latter sequence,
% the selection algorithm will also have difficulty deciding which \NLST to mutate.

% In addition,
% the performance of \AFLNet selection algorithm can also be affected by its empirical state evaluation function.
% Such function is not supported by statistic reasoning or previous studies.\MYTODO{Thuan: Any other issues in the current \AFLNet score function?}

\subsection{Algorithm} \label{sec: algorithm}
% \MYTODO{Thuan: we probably should reorganise this (sub)section. It focuses a lot on the tree state model that assists state selection, but very little on the UCT function from MCTS.
% Maybe we can merge them into one? The tricky task is to make it consistent withe the Motivating Example after merging them}

Observing the first weakness of \AFLNet's algorithm,
\AFLNetLegion proposes to
% the following two improvements:
% 1) 
assist state selection with a more fine-grained 
% request sequence partition with a 
tree model illustrated by \cref{fig: AFLNetLegion tree}.
% 2) extends \Legion's principled state selection algorithm to stateful protocol fuzzing.
% \begin{figure*}[ht!]
%     \centering
%     \includegraphics[width=\linewidth]{Figures/AFLNetLegionModel}
%     % \input{Figures/AFLNetModel.tex}
%     \vspace*{-.3cm}
%     \caption{\AFLNet's state machine model of \cref{fig: RequestSeqs}}
%     \label{fig: AFLNet state machine}
% \end{figure*}
% \subsubsection{Tree Model}
% \cref{fig: AFLNetLegion tree} illustrates \AFLNetLegion'
The state model construction is the same as \AFLNet,
except every node is identified with a unique response code sequence from the root to itself.
% The tree root is again a dummy node representing the initial state of the server,
% while every child node corresponds to a subsequent response code of the latest command.
% Every node is identified with a unique response code sequence from the root to itself.
Each also stores the same statistics and request sequences
% of its response code sequence 
as \AFLNet.
% , observed passing through that node so far.

The tree model allows a more fine-grained server state modelling,
by unfolding the states from \AFLNet's model into tree branches.
% In this way,
% it can distinguish the corresponding internal states of two nodes not only by their own response codes,
% but also by all previous response codes in the same session.
For example,
the interesting \NLST was sent from node \texttt{250''},
which is clearly distinguished from other nodes with the same response code but different prefixes
(e.g. \texttt{250}, \texttt{250'}, \texttt{250'''}),
so that they will be evaluated and selected independently.
% This ensures the model will 
% no longer request sequence 2 or similar interesting sequences with the same prefix into the node.
Consequently,
no noise request sequence will 
undermine the actual statistics of each node or complicate its state selection.
\cref{subsec: tree nodes} and \cref{subsec: tree construction} explain the detail of this model.
% \subsection{A principled algorithm} \label{sec: aflnetlegion - overview - a principled algorithm}

% \subsubsection{Principle algorithm}
% Inspired by the success of \cref{sec: aflnetlegion - background - monte carlo tree search and legion},
% % Based on the observation from \cref{sec: aflnetlegion - overview},
% We extends \Legion's principled state selection algorithm to stateful protocol fuzzing.
% Specifically, this algorithm, termed \AFLNetLegion,
% replaces the empirical state evaluation formula in \AFLNet
% with a principled UCT function.
% % This section describes the key modelling decisions, respectively 
% % the tree nodes (\cref{subsec: tree nodes}),
% % the tree construction (\cref{subsec: tree construction}),
% % and the selection policy of nodes (\cref{subsec: selection policy}),
% % to adopt the framework.

% \cref{subsec: selection policy} explains the detail of this algorithm.

\subsection{Tree Nodes} \label{subsec: tree nodes}
Tree nodes are designed to allow \AFLNetLegion to fuzz the server from past server states preferred by the its MCTS-based algorithm.
Each state of the server under test is identified by a sequence of response codes received during fuzzing.

Recall from the example in \cref{sec: algorithm},
the root of the tree represents the initial state of the server (denoted by a dummy code `0'), 
and every child node corresponds to a response code received after its parent code.
Each tree node stores \textit{interesting} request sequences that can reproduce the corresponding server state.
A request is considered interesting to a node if it finds a new child of the node.
This is because,
under the MCTS paradigm,
we consider all requests that produce the same response code from the same server state
as an identical action,
and only save (the first) one of them to avoid duplication.
To facilitate state selection,
each node also stores its selection count and discovery count,
% of the node and each request sequence saved in the node.
which respectively record the number of times it has been selected
and the number of new child it has found.
The same two statistics of each request sequence are also saved to support their selection.

Inherited from \Legion,
\AFLNetLegion's nodes come in three types for different purposes.
We use \white nodes to represent the server states where we can launch fuzzing (e.g. right after executing \texttt{STOR sample.text}
% and before it receives \texttt{NLST *} 
in \cref{lst: request sequence 2}).
In this way, we can fuzz interesting requests (e.g. \texttt{NLST *}) from the same server state as they were previously sent.
A \golden node is attached to each \white node,
representing the option of launching fuzzing from the latter.
\cref{fig: aflnetlegion phases} draws \white nodes and \golden nodes respectively as circles and squares.
In some cases,
one request can trigger multiple response codes in a row,
in which case we can only launch fuzzing from the server state represented by the last code in them.
The last code of them is represented by a \white node
while the others are represented by \black nodes.
The latter is omitted from \cref{fig: aflnetlegion phases} as they do not affect fuzzing.

Note that the \red node from \Legion is not applicable in \AFLNetLegion,
since symbolic execution is not available in the scope of this paper.
Consequently,
we can never know if all children of a node have been found,
which highlights the necessity of a principled selection algorithm
(\cref{subsec: selection policy})
that can upper-bound the expected loss in coverage due to not selecting the best server state.

\subsection{Tree Construction} \label{subsec: tree construction}
\begin{figure*}[ht!]
    \centering
    \input{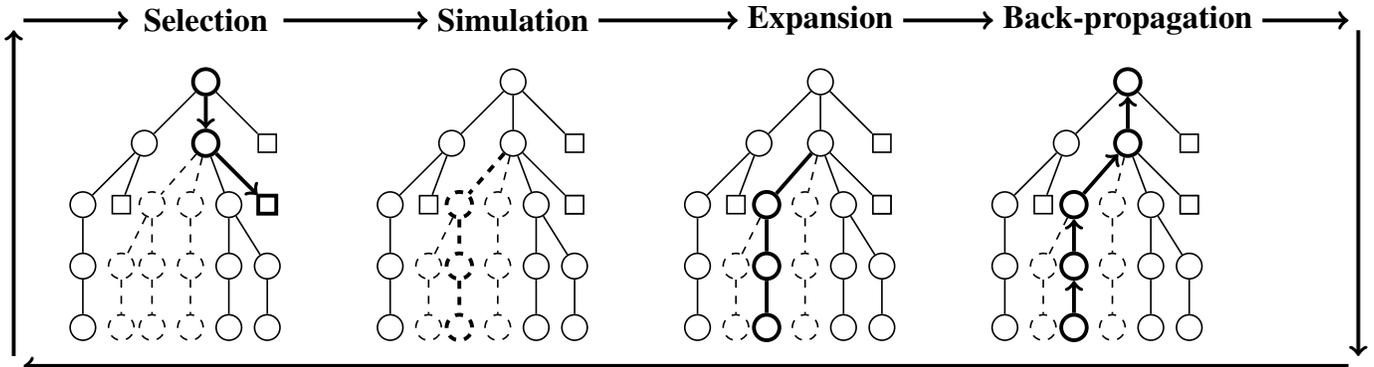}
    \vspace*{-.3cm}
    \caption{The four stages of (each iteration of) \AFLNetLegion's MCTS algorithm.}
    \label{fig: aflnetlegion phases}
  \end{figure*}

  \cref{fig: aflnetlegion phases} illustrates how \AFLNetLegion
explores the tree-structured search space in its variation of MCTS:
Each iteration of the search algorithm proceeds in the same four steps as \Legion with minor differences tailored to stateful protocol fuzzing.
In \cref{fig: aflnetlegion phases}, bold lines highlight the actions in each stage.
Solid thin lines and nodes represent response sequence and server states that have been covered and integrated into the tree.
Dashed lines and nodes draw the undiscovered response sequences and server states.
The four steps execute in the following order:

% \MYTODO{Make sure they are consistent with the new narrative in Intro}
\bparagraph{Selection.}
% \MYTODO{
% Question: Is it ok to also mention request sequence selection here? I did it because the experiment needs it (e.g. \legionuu vs. \legionur)}
It consists of two sub-steps in a row: 
Tree node selection and seed request sequence selection.
The tree node selection descends from the root node 
by recursively applying the \textit{selection policy}, until it reaches a \golden node.
Then it applies the selection policy on 
all the request sequences stored in the tree node to find a seed for fuzzing.
We leave the rule of the selection policy to \cref{subsec: selection policy}.

\bparagraph{Simulation.}
% Like \AFLNet, \AFLNetLegion adopts \AFL for fuzzing.
% It explores different behaviour of the server after the selected state.
It replays the request to restore the server state corresponds to the parent \white node (e.g. Node $250''$ in \cref{fig: AFLNetLegion tree}), and then fuzzes the following requests (e.g. \texttt{NLST *}).
% Among the three seed request sequence sections marked by \AFLNet
% (discussed in \cref{sec: aflnetlegion - background}),
% \AFLNetLegion replays the prefix section
% and mutates both the candidate and the suffix sections.
% This design decision follows the random policy of the MCTS paradigm.

\bparagraph{Expansion.}
It records all newly discovered response sequences and their request sequences into the tree.
Inherited from \Legion,
\AFLNetLegion maps each observed response code sequence to a path of tree nodes.
When a new response code subsequence is found,
it adds each of the new code as a new child node of its predecessor,
and saves the corresponding request sequence to the parent of the first new node
and every new descendants.
% It also marks the three sections of the new request sequence when saving it to each node
% so that the sequence can be reused in the future simulation stages.
As mentioned,
when a single request triggers multiple response codes in a row,
only the last code of such subsequence will be created as a \white node
while all its ancestors of the same subsequence (if any) will be added as \black nodes.
Again, a \golden node will be attached to each \white node.

\bparagraph{Backpropagation.}
It updates the statistics recorded in the tree to assist future decision making.
For each mutant executed in the simulation stage,
\AFLNetLegion increases the \textit{past selection count} of 
each node in the selection path by one to
discourage selecting the same nodes too often.
For each new response code sequence observed,
it increases the \textit{past discovery count}
of the parent of the first new node and all subsequent new nodes,
to encourage selecting them in the future.
The usage of these two counts are discussed in \cref{subsec: selection policy}.

% Similarly,
% for each new path selected or found by fuzzing the selected input,
% this step respectively increases the selection count or reward count of that input by one.

% point the readers to related work for a detailed comparison between aflnet aflnetlegion and legion

\subsection{Selection Policy} \label{subsec: selection policy}
To mitigate the second weakness of \AFLNet,
\AFLNetLegion's selection policy takes charge of 
choosing tree nodes and seed request sequences based on their statistics
in the selection stage of MCTS.

It estimates the productivity of a server state regarding finding more uncover states.
It relies on the same two statistics as \AFLNet,
namely,
\textit{past discovery count} and \textit{past selection count}.
Again, the former represents the number of response code sequences found
under the target program state,
more discovery indicates higher productivity of the state.
The latter stands for the number of times the target state has been selected for fuzzing,
less selection implies higher uncertainty about the state.
Together, they attempt to
balance exploitation of the states that appear to be most productivity based on past experience, 
against exploration of less-well understood parts of the protocol
whose productivity (response code sequence discovery) are less certain.

\begin{equation}
% \begin{gather*}
  \UCT(N) =  
  \begin{cases}
    \infty  & S=0\\
    \frac{D}{S} + \rho\sqrt{ \frac{2\ln{P_S}}{S} } & S > 0
  \end{cases}
% \end{gather*}
\label{eqn: UCT}
\end{equation}

% {\color{red} move the above to background?}

% The reason behind this heuristic is explained in detail in \Legion.
Different from \AFLNet (which relies on intuitive formulas),
\AFLNetLegion inherites the UCT function (\cref{eqn: UCT}) from \Legion to evaluate both tree nodes and request sequences:
For node evaluation,
$D$, $S$, and $P_S$ respectively denotes 
\textit{past discovery count} of the node,
\textit{past selection count} of the node,
and \textit{past selection count} of the nodes parent.
For seed evaluation,
$D$, $S$, and $P_S$ respectively denotes 
\textit{past discovery count} of the seed,
\textit{past selection count} of the seed,
and \textit{past selection count} of the seed's belonging simulation node.
The discovery count records 
the number of new response sequences found by selecting that state/seed,
the selection count records the number of times the state/seed has been selected for fuzzing.

With both main weaknesses mitigated,
the next section will evaluate the performance of \AFLNetLegion.

\section{Evaluation} \label{sec: aflnetlegion - evaluation}
This section addresses two research questions:
% The experiments serve for one of the following two logic:
% (A) A better search algorithm can improve the coverage performance, but not much
% (B) AFLNetLegion comes with a fine-grained model and an input selection algorithm
% Its selection algorithm works for some particular cases, but its overall performance is limited
% Its fine-grained model contributes more to the coverage performance improvement

\bparagraph{
(1) Is there any statistical evidence to support that
 \AFLNetLegion's state selection algorithm may be better than \AFLNet?} 
 % (1) Should we expect \AFLNetLegion's model and algorithm to perform better than \AFLNet?
To answer this question,
we conduct a detailed \emph{case study} to analyse the coverage performance of \AFLNet and \AFLNetLegion 
on ten sample code blocks carefully selected from the \glob library of protocol \proftpd, as an extension to the motivating example in \cref{subsec: motivating example}.

\bparagraph{
(2) How do different modelling and selection algorithms impact \textit{overall} coverage performance?}
% \bparagraph{Overall comparison.}
To answer this question, 
we compare the \emph{overall coverage performance} of
$3$ combinations of state and seed selection algorithms based on \AFLNetLegion against each of the best-performing \AFLNet algorithm from the 6 benchmark subjects from \cref{sec: aflnet - preliminary}, following the same experiment setup.

% The case study in \cref{aflnetlegion - evaluation - case study} and 
% the overall performance comparison in \cref{aflnetlegion - evaluation - overall performance} 
% respectively answer these two questions.

% \MYTODO{highlight the research question, instead of the following two parts}
% \bparagraph{Case study.}
% To answer the first question,
% we conduct a detailed study to analyse the coverage performance of \AFLNet and \AFLNetLegion 
% on ten sample code blocks carefully selected from the \texttt{glob} library of protocol \proftpd.

% \bparagraph{Overall comparison.}
% To answer the second question, 
% we compare the overall coverage performance of
% $6$ combinations of server models, server state selection algorithms, and seed selection algorithms.

\subsection{Case Study} \label{aflnetlegion - evaluation - case study}
% To answer the first question, 
% we compare and analyse the coverage performance of \AFLNet and \AFLNetLegion on 
% ten code blocks from the \texttt{glob} library of protocol \proftpd.
% We choose the ten cases to investigate not because they are the only cases where 
% \AFLNetLegion outperforms \AFLNet.
% In fact,
% \AFLNetLegion exhibits better performance than \AFLNet on \proftpd as a whole 
% and on other benchmarking subjects as well
% (see \cref{aflnetlegion - evaluation - overall performance}).
% We select these ten cases because they are consistent with 
% the example used in \cref{sec: aflnetlegion - overview},
% and because they are representative enough to show why
% we can expect \AFLNetLegion to perform better.
% rare but interesting
% \textit{glob} via \NLST queries .
% This functionality is implemented in the ,
% it expands wildcard patterns (\texttt{*}, \texttt{?}, or \texttt{[]}) 
% into a list of pathnames matching the pattern.

Analysing the coverage of the \glob library can tell us if 
\AFLNetLegion's state selection algorithm is capable of improving coverage performance.
Although the library appears easy to \textit{reach},
there are many special cases that are difficult to \textit{fully cover}.
The former ensures most algorithms can generate some coverage statistics for comparison and analysis,
while the coverage performance of the latter cases tells how efficient/effective the algorithm is.
Recall the motivating example,
the first three lines from \cref{lst: request sequence 1} is a minimum test case to reach this model without testing many lines in the library.
% is sending a \NLST query with any wildcard parameter right after logging in as a valid user,
% like the \NLST command in the sequence 1 in \cref{fig: RequestSeqs}.
% However,
% such a simple request sequence will not cover many lines in the module, 
% as it will fail to match anything, given no file has been created.
Alternatively,
\cref{lst: request sequence 2}
is able to test \glob's functionality of expanding the wildcard by creating some files beforehand.
% trigger a successful globbing.
% like the right request sequence in \cref{fig: RequestSeqs}.
Similarly,
many other code blocks also correspond to particular cases,
such as
% \MYTODO{1 -> Case 1}
% \FIXED{}
% Strange alignment with the following code
% \begin{enumerate}
%     \item[Case 1] Failing to allocate more memory
%     \item[Case 2] Globbing a file without path prefix
%     \item[Case 3] Globbing a file with a path prefix
%     \item[Case 4] Having a wildcard in a directory name
%     \item[Case 5] Globbing with metacharacter "\texttt{*}" or "\texttt{?}"
%     \item[Case 6] Globbing with metacharacter "\texttt{[]}"
%     \item[Case 7] Escaping a wildcard metacharacter
%     \item[Case 8] Successfully expanding a directory with a metacharacter
%     \item[Case 9] Globbing a directory that contains "\texttt{\textbackslash\textbackslash}"
%     \item[Case 10] Successfully finding at least 1 match 
% \end{enumerate}

\begin{itemize}
    \item Case 1: Failing to allocate more memory
    \item Case 2: Globbing a file without path prefix
    \item Case 3: Globbing a file with a path prefix
    \item Case 4: Having a wildcard in a directory name
    \item Case 5: Globbing with metacharacter "\texttt{*}" or "\texttt{?}"
    \item Case 6: Globbing with metacharacter "\texttt{[]}"
    \item Case 7: Escaping a wildcard metacharacter
    \item Case 8: Successfully expanding a directory with a metacharacter
    \item Case 9: Globbing a directory that contains "\texttt{\textbackslash\textbackslash}"
    \item Case 10: Successfully finding at least 1 match 
\end{itemize}

The case study will compare 12 hours performance of the selection algorithms of \AFLNet and \AFLNetLegion on the ten cases above.

\subsection{Overall Evaluation} \label{aflnetlegion - evaluation - overall performance}
While the case study highlights performance on particular interesting code blocks,
the overall evaluation zooms out to the big picture.
It investigates the overall impact of state selection algorithm
on different protocol implementations.
% We evaluate four algorithms on the same six stateful protocol implementations provided by \ProFuzzBench.
% The four selection algorithms are summarised in \cref{tbl: overall algorithm}.
% Two of them are from \AFLNet, the other two are based on \AFLNetLegion,
% For \AFLNet,
% the round robin approach selects each server state and request sequence in turns,
% the favour approach attempts to balance exploration and exploitation with an empirical function.
% For \AFLNetLegion,
% the UCT approach selects server states and request sequences
% with the highest score computed with \cref{eqn: UCT},
% the random approach enforces a uniform random selection.
% As an controlled experiments,
% the random approach still descends the tree as a normal selection stage of MCTS,
% resulting in selecting the top tree nodes more often.

% The six stateful protocol implementations come from five categories, including 
% two \textit{FTP} servers (\textit{LightFTP} and \textit{ProFTPD}), 
% one \textit{SSH} server (\textit{OpenSSH}),
% one \textit{TLS} server (\textit{OpenSSL}),
% one \textit{SMTP} server (\textit{Exim}), and 
% one \textit{RTSP} server (\textit{Live555}).

% \begin{table*}[ht!]
%     \centering
%     \resizebox{\linewidth}{!}{\input{Tables/OverallAlgorithms.tex}}
%     \caption{Algorithms in the overall evaluation}
%     \label{tbl: overall algorithm}
% \end{table*}

\subsection{Experiment Setup} \label{subsec: aflnetlegion - evaluation - experiment setup}

% We put the following effort to produce trustworthy results
% according to the most thorough evaluation guidelines to our knowledge \cite{klees2018evaluating}.

We used the same experimental setup as described in \cref{sec: aflnet - preliminary} for this evaluation. Additional details follow.

\iparagraph{Case Study-Specific Setup.}
%\iparagraph{Benchmarking programs.}
Desipte \AFLNetLegion exhibits better overall performance than \AFLNet in \proftpd,
the case study chooses only ten code blocks to investigate because
their consistency with the example in \cref{subsec: motivating example},
and their statistical evidence are typical to show why \AFLNetLegion can perform better.
We repeated each experiments for $50$ trials to support a statistically meaningful t-test evaluation.

\iparagraph{Algorithms and Models.}
In addition to the best-performing algorithm of \AFLNet on each benchmark program from \cref{sec: aflnet - preliminary},
we also add three new algorithms based on \AFLNetLegion.
\AFLNetLegion uses the same backend as \AFLNet,
all differences come from their model and algorithm discussed in the previous sections.
Its default algorithm \legionuu selects both states and seeds with the highest UCT scores. An alternative, \legionur, uses UCT for state selection and selects seeds randomly. The third algorithm \legionrr select both states and seeds randomly.
% We did not include other fuzzers in the evaluation
% because so far \AFLNet is the most suitable backend to adopt \Legion's algorithm,
% and to control the implementation in these experiments.
% because so far \AFLNet is the only fuzzer that is capable of 
% fuzzing external servers with sequences of inputs.
The source code of 
\AFLNet \footnote{https://github.com/Alan32Liu/aflnet/tree/SANER2022} and 
\AFLNetLegion
\footnote{https://github.com/Alan32Liu/AFLNet\_Legion/tree/SANER2022} 
used in this paper are publicly available.
% In the case study,
% we use the \favour algorithm of \AFLNet as the baseline to compare against
% the \mctsf algorithm of \AFLNetLegion,
% as both algorithms are considered and proven to be the best of each fuzzer.
% These two algorithms are respectively explained in
% \cref{sec: aflnetlegion - background - aflnet} and \cref{sec: aflnetlegion - algorithm}.
% In the overall comparison,
% we examined the coverage performance of $2$ algorithms from \AFLNet 
% (\aflnetbb, and \aflnetff), 
% and $3$ from \AFLNetLegion (\legionur, \legionuu).
% More detail of these algorithms are explained in
% \cref{aflnetlegion - evaluation - overall performance}.
\subsection{Experiment Results}
\subsubsection{Case Study} \label{sec: aflnetlegin - experiment result - case study}
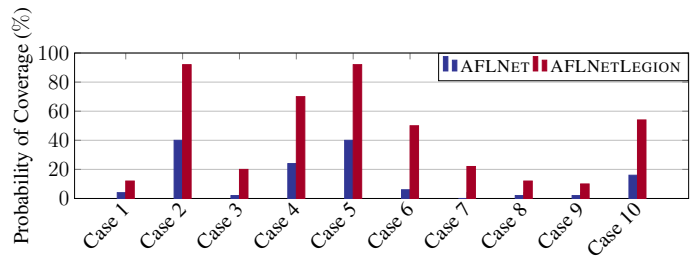
\begin{figure}[ht!]
    \centering
    % \definecolor{my_blue}{hex}{0.12156862745098039, 0.4666666666666667, 0.7058823529411765}
% \definecolor{my_red}{hex}{0.8392156862745098, 0.15294117647058825, 0.1568627450980392}
\definecolor{my_blue}{HTML}{313695}
\definecolor{my_red}{HTML}{a50026}

\begin{adjustbox}{width=0.5\textwidth}
\begin{tikzpicture}[ybar]
    \begin{axis}[
        width=16cm,
        height=5cm,
        bar width=2mm,
        label style={font=\Large},
        tick label style={font=\Large},
        % y ticks style and label
        % axis y line*=left,
        ybar,
        ylabel={Probability of Coverage ($\%$)},
        ymin=0, ymax=100,
        ytick={0,20,...,100},
        ymajorgrids,
        % x axis ticks and style
        symbolic x coords={
            Case 1,
            Case 2,
            Case 3,
            Case 4,
            Case 5,
            Case 6,
            Case 7,
            Case 8,
            Case 9,
            Case 10,
        },
        xtick=data,
        xtick align=inside,
        x tick label style = {xshift=2.5, yshift=-2.5, rotate=45, anchor=east},
        % legend
        legend style = {at={(1,1), font=\tiny, anchor = east, font=\large},
        legend columns=-1,
        % /tikz/every even column/.append style={column sep=1em}
        }
    ]
        % \draw[color=red,fill=red!80,bar width=6pt]
        % plot coordinates{
        \addplot[color=my_blue, fill=my_blue, bar shift=-1mm] coordinates {
            (Case 1, 4) 
            (Case 2, 40) 
            (Case 3, 2) 
            (Case 4, 24) 
            (Case 5, 40) 
            (Case 6, 6) 
            (Case 7, 0) 
            (Case 8, 2) 
            (Case 9, 2) 
            (Case 10, 16)
        };
        % \draw[color=red!50,fill=red!20,bar width=4pt,bar shift=3pt]
        % plot coordinates{
        \addplot[color=my_red, fill=my_red, bar shift=1mm] coordinates {
            (Case 1, 12)
            (Case 2, 92)
            (Case 3, 20)
            (Case 4, 70)
            (Case 5, 92)
            (Case 6, 50)
            (Case 7, 22)
            (Case 8, 12)
            (Case 9, 10)
            (Case 10, 54)
        };
        \legend{\AFLNet, \AFLNetLegion}
    \end{axis}
\end{tikzpicture}
\end{adjustbox}
    % \vspace*{-.3cm}
    \caption{Coverage performance on the \glob library 
            %  {\color{red} \FIXME{
            %      Resize?;
            %      Suitable fonts (xy label, legend, etc.);
            %      Reorder?;
            %      More descriptive x label names?;
            %      Show \AFLNet is how many times better than \AFLNEt?}
            %      }
            }
    \label{fig: globbing result}
\end{figure}

\cref{fig: globbing result} compares the coverage performance of \AFLNet and \AFLNetLegion
on the ten representative cases in \glob library.
For both algorithms,
we count the number of trials that each case is covered
and use the percentage value (out of $50$) to represent their likelihood of covering
that case in an average trial.
For example,
% out of $50$ trials of each algorithm,
$1$ trial of \AFLNet and $10$ trials of \AFLNetLegion covered the third case,
hence the probabilities are respectively $2\%$ and $20\%$.

% \begin{table*}[ht!]
%     \centering
%     % \input{case study - input seqs.tex}
%     \resizebox{\linewidth}{!}{\input{Tables/GlobbingResult.tex}}
%     \caption{Coverage performance on the glob library
%              {\color{red} \FIXME{Better in a bar chart (i.e. \cref{fig: globbing result})?}}
%             }
%     \label{tbl: globbing result}
% \end{table*}

\AFLNetLegion achieves a better overall coverage performance on the whole library
because its higher probabilities in covering all ten interesting cases.
For each case,
\AFLNetLegion at least doubled \AFLNet's coverage probability.
These cases are difficult to cover either because of their
a) \emph{non-trivial prerequisites} or b) \emph{rareness}.

Covering cases with non-trivial prerequisites requires
the ability to accurately select server states that satisfy the requisites to launch fuzzing.
As discuss in \cref{sec: algorithm},
\AFLNetLegion fulfils this requirement via
unfolding the compact state machine into a tree model,
%  each request is sent
where each node is distinguished by all the response codes received in the current session.
In this way,
it reduces noise in state selection
(i.e., states that share the same response code but did not fulfil the prerequisites),
allows a more accurate measurement of past reward of each state,
% (interesting response sequences discovery),
and constitutes a reliable estimation of the potential of each server state.
For example, case 8 requires globbing an existing directory,
which will be missed if a fuzzer's algorithm failed to select a server state that has some directories created beforehand.
Mixing such states with noises will
largely lower the chance of covering this case.
As a result,
an average \AFLNet trial has merely $2\%$ probability to covers this case,
while \AFLNetLegion increases this probability by $6$ times.

Covering rare cases requires the ability to balance selecting states and request sequence
that appear to be most rewarding
and ones where rewards are less certain.
Such ability allows intentionally generating enough mutants from the states that do not seem productive at first.
For example,
case 7 requires the globbing pattern to
contain a "\texttt{[}" followed by a "\texttt{]}" some characters later.
Without knowing this specific grammar,
\AFLNetLegion was able to cover this line more than once in an average trial.
As a comparison,
although \AFLNet's heuristic uses the same statistics (i.e. discovery and selection count),
none of its $50$ trials was able to cover this case.

Both abilities constitute \AFLNetLegion's advantage in server state estimation and selection.
With an accurate and reliable estimation of the complexity of the \glob library,
\AFLNetLegion on average generated over $14$ times more test cases for it than \AFLNet.
% \AFLNetLegion on average generated more than $289$ test cases for it.
% while \AFLNet only reached the library for approximately $20$ times.
Recall \cref{subsec: motivating example} once gave an example to test \glob library with \NLST requests.
\AFLNetLegion on average generated over $15$ times more \NLST requests than \AFLNet,
despite that the total number of all requests generated by both algorithms are of the same magnitude.
\AFLNetLegion intentionally focused on \NLST commands because its algorithm
% can find the states that satisfy the interesting prerequisites and also capable to 
can identify the existence of interesting cases hidden inside the library,
and notice mutants of \NLST requests can cover them.
% and intentionally generate more than $10$ times test cases for them.
% Note that the total number of requests generated by both algorithms are of the same magnitude.

% because it is more capable of covering
% We can expect a better coverage performance from \AFLNetLegion,
This experiment proves that
we can \emph{expect} \AFLNetLegion to perform better with its algorithm.
In \cref{aflnetlegion - evaluation - overall performance},
we will compare this algorithm against various alternatives
regarding the impact on the overall coverage performance in 24-hour experiments.

\subsubsection{Overall Comparison}
\label{sec: aflnetlegin - experiment result - overall comparison}

\begin{figure*}[ht!]
    \centering
    \begin{subfigure}[b]{0.32\textwidth}
        \centering
        \includegraphics[width=\textwidth]{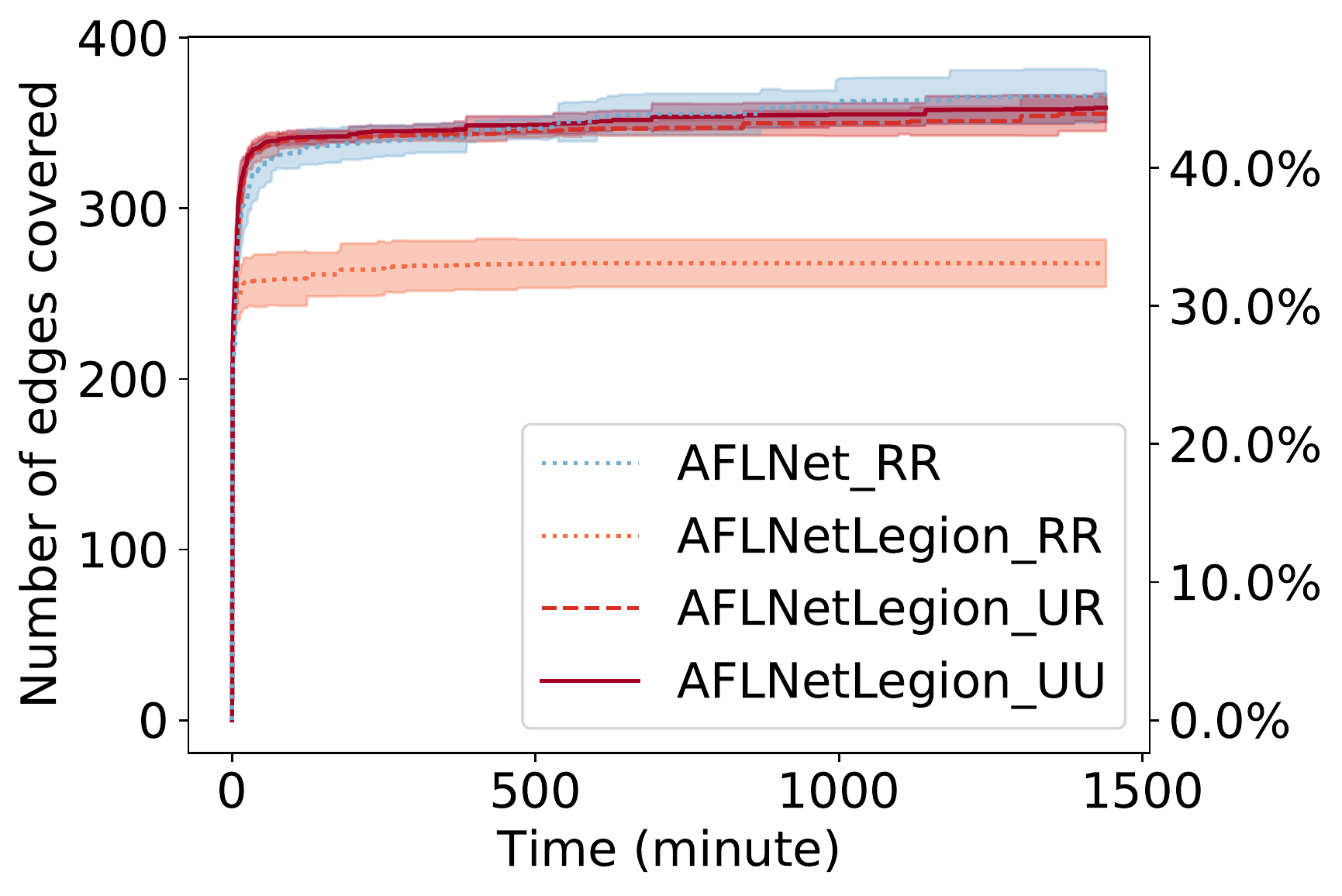}
        \caption{\texttt{LightFTP} branch coverage}
        \label{fig: overall lightftp branch coverage}
    \end{subfigure}
    \hfill
    \begin{subfigure}[b]{0.32\textwidth}
        \centering
        \includegraphics[width=\textwidth]{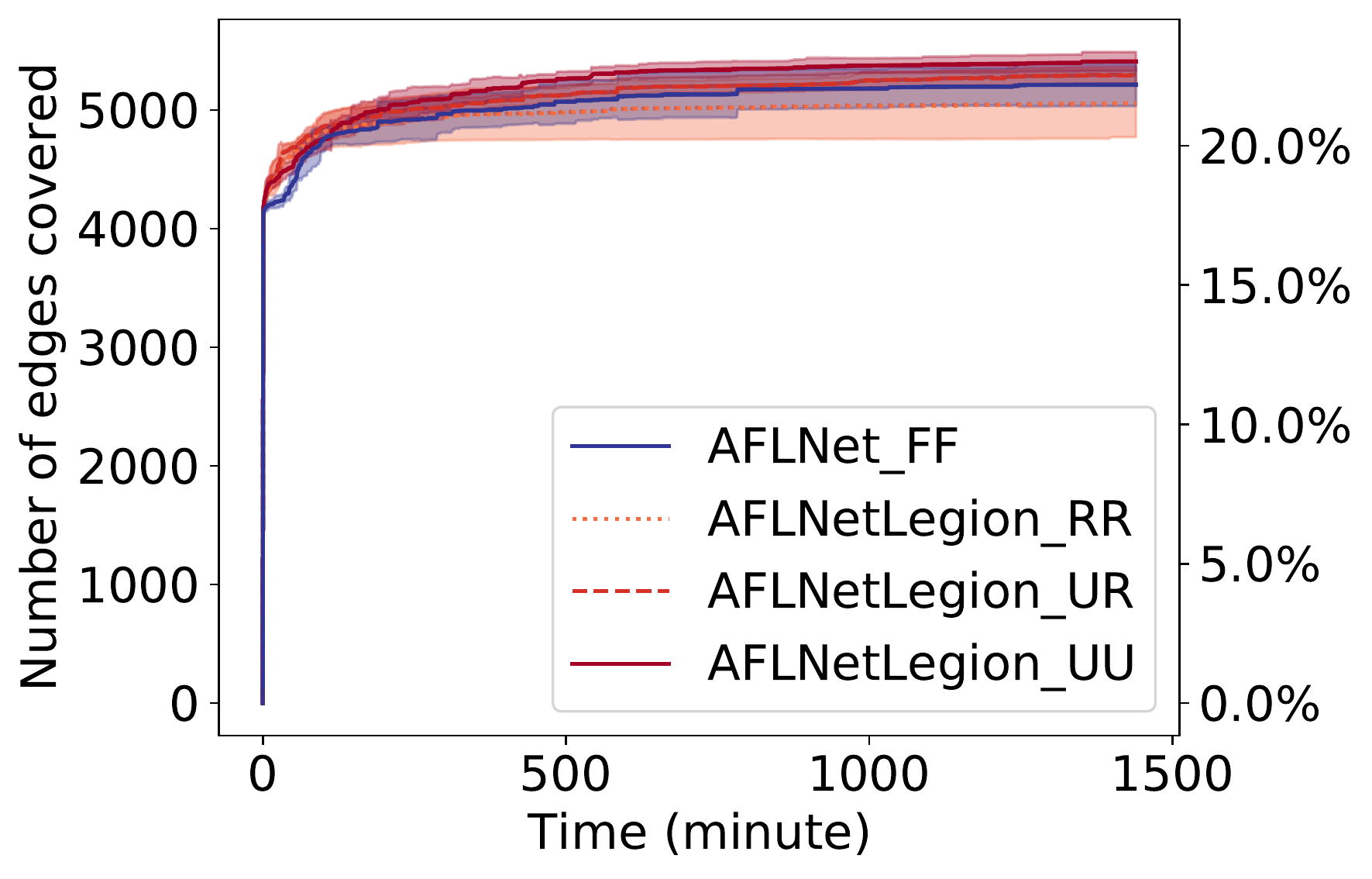}
        \caption{\texttt{ProFTPD} branch coverage}
        \label{fig: overall proftpd branch coverage}
    \end{subfigure}
    \hfill
    \begin{subfigure}[b]{0.32\textwidth}
        \centering
        \includegraphics[width=\textwidth]{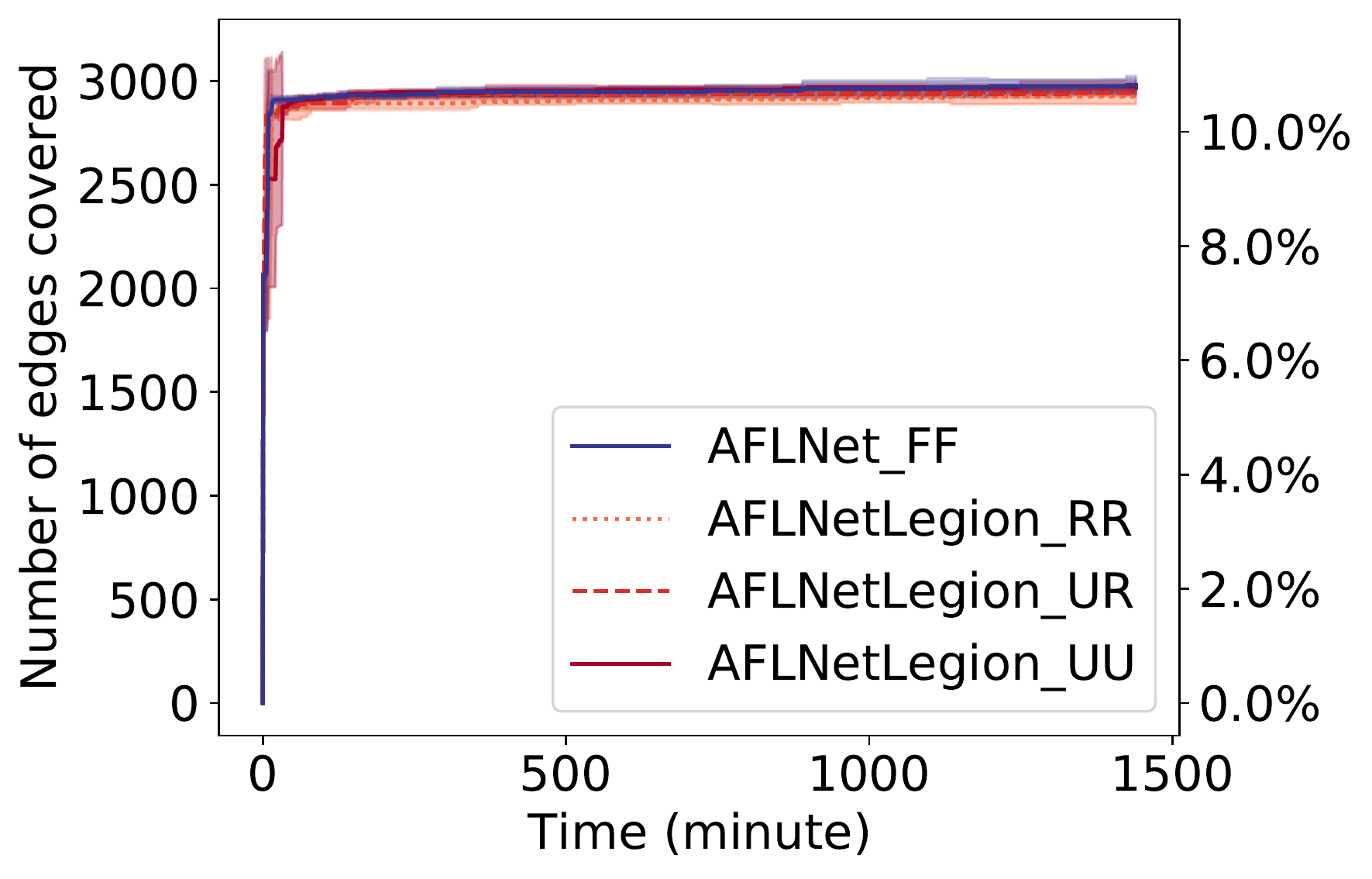}
        \caption{\texttt{Exim} branch coverage}
        \label{fig: overall exim branch coverage}
    \end{subfigure}
    % \vskip\baselineskip
    % \begin{subfigure}[b]{0.32\textwidth}
    %     \centering
    %     \includegraphics[width=\textwidth]{Figures/lightftp/l_random_best.pdf}
    %     \caption{\texttt{LightFTP} line coverage}
    %     \label{fig: overall lightftp line coverage}
    % \end{subfigure}
    % \hfill
    % \begin{subfigure}[b]{0.32\textwidth}
    %     \centering
    %     \includegraphics[width=\textwidth]{Figures/proftpd/l_favour_best.pdf}
    %     \caption{\texttt{ProFTPD} line coverage}
    %     \label{fig: overall proftpd line coverage}
    % \end{subfigure}
    % \hfill
    % \begin{subfigure}[b]{0.32\textwidth}
    %     \centering
    %     \includegraphics[width=\textwidth]{Figures/exim/l_favour_best.pdf}
    %     \caption{\texttt{Exim} line coverage}
    %     \label{fig: overall exim line coverage}
    % \end{subfigure}
\end{figure*}
\begin{figure*}[ht!]\ContinuedFloat
    \begin{subfigure}[b]{0.32\textwidth}
        \centering
        \includegraphics[width=\textwidth]{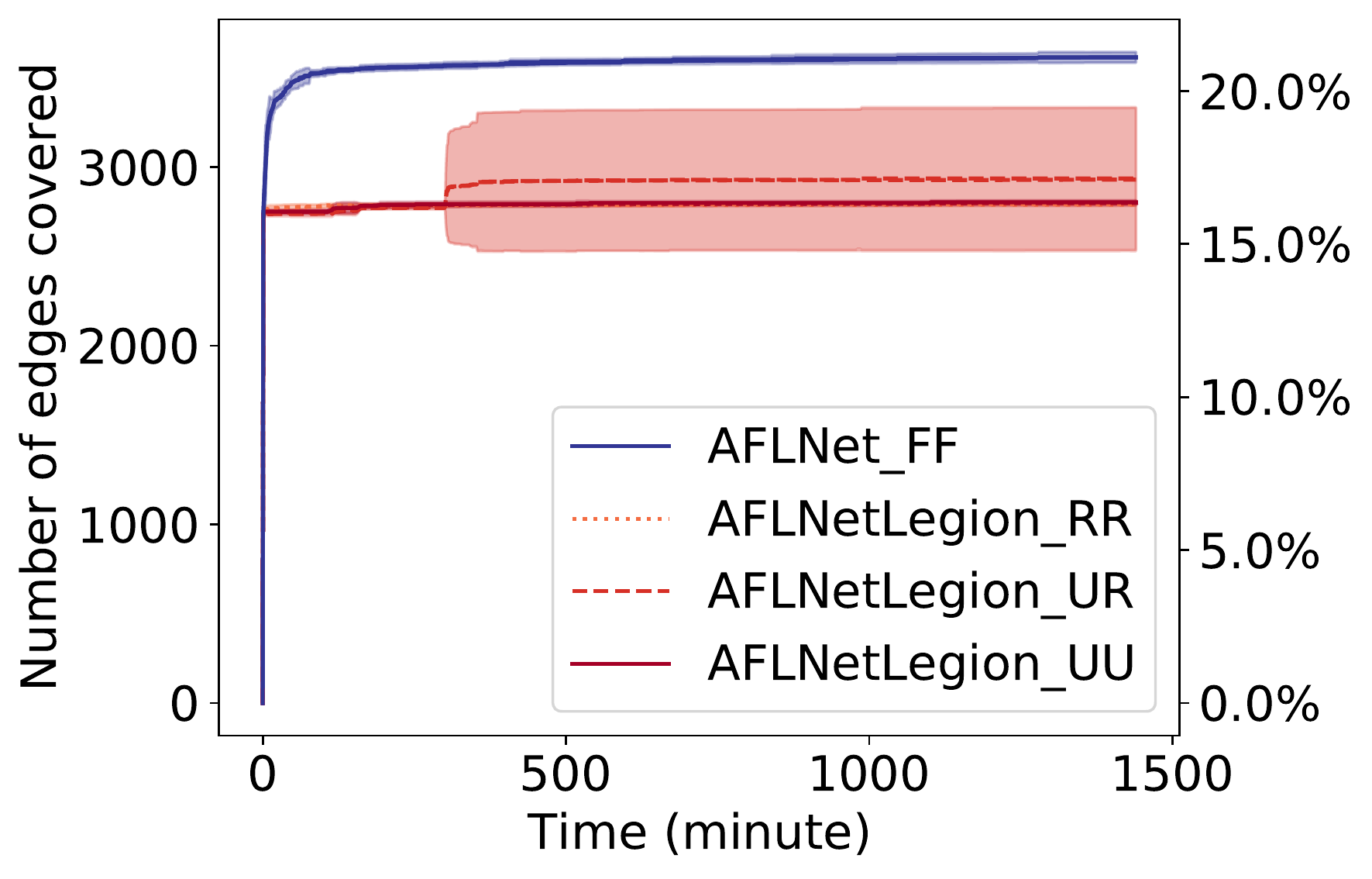}
        \caption{\texttt{OpenSSH} branch coverage}
        \label{fig: overall openssh branch coverage}
    \end{subfigure}
    \hfill
    \begin{subfigure}[b]{0.32\textwidth}
        \centering
        \includegraphics[width=\textwidth]{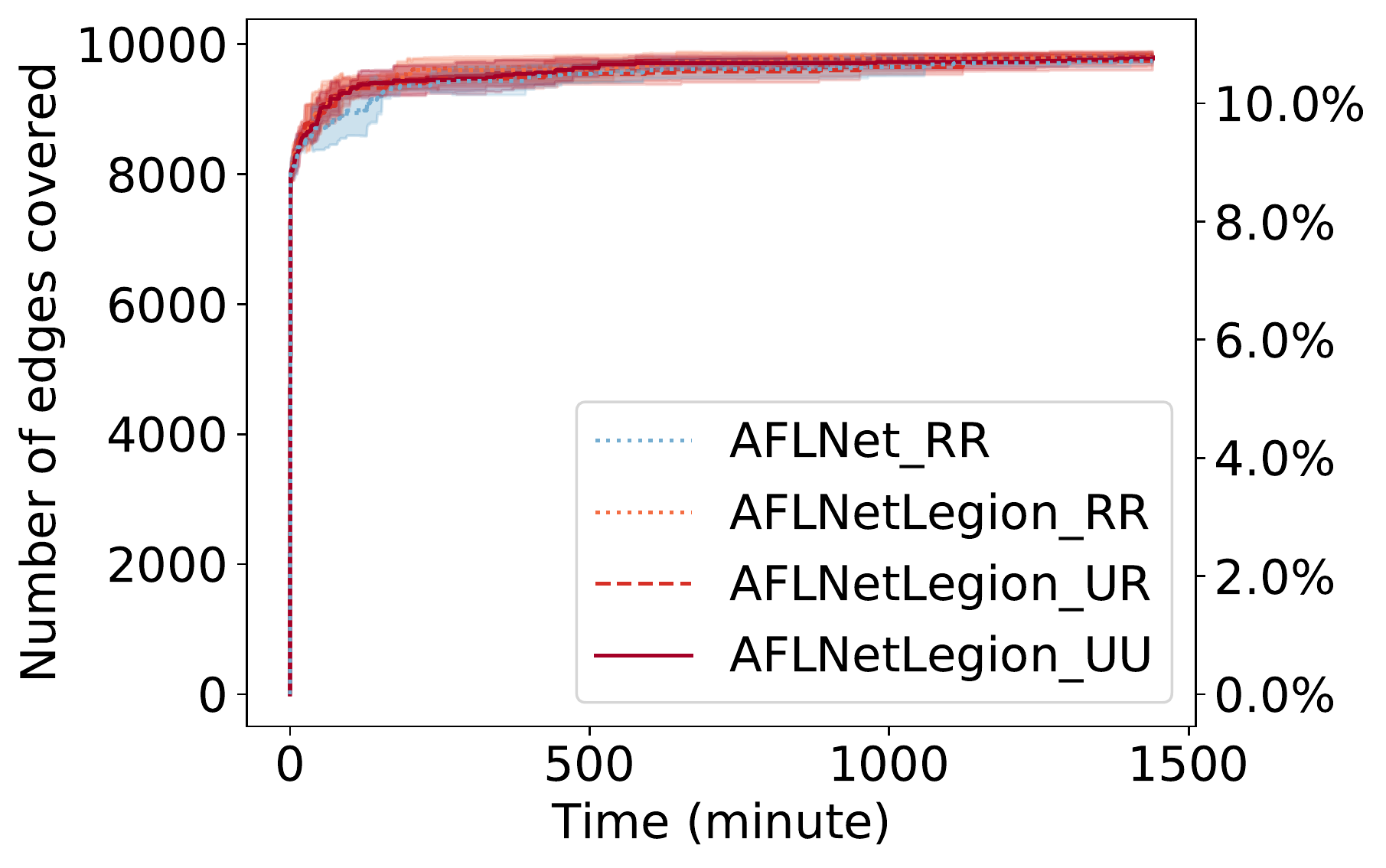}
        \caption{\texttt{OpenSSL} branch coverage}
        \label{fig: overall openssl branch coverage}
    \end{subfigure}
    \hfill
    \begin{subfigure}[b]{0.32\textwidth}
        \centering
        \includegraphics[width=\textwidth]{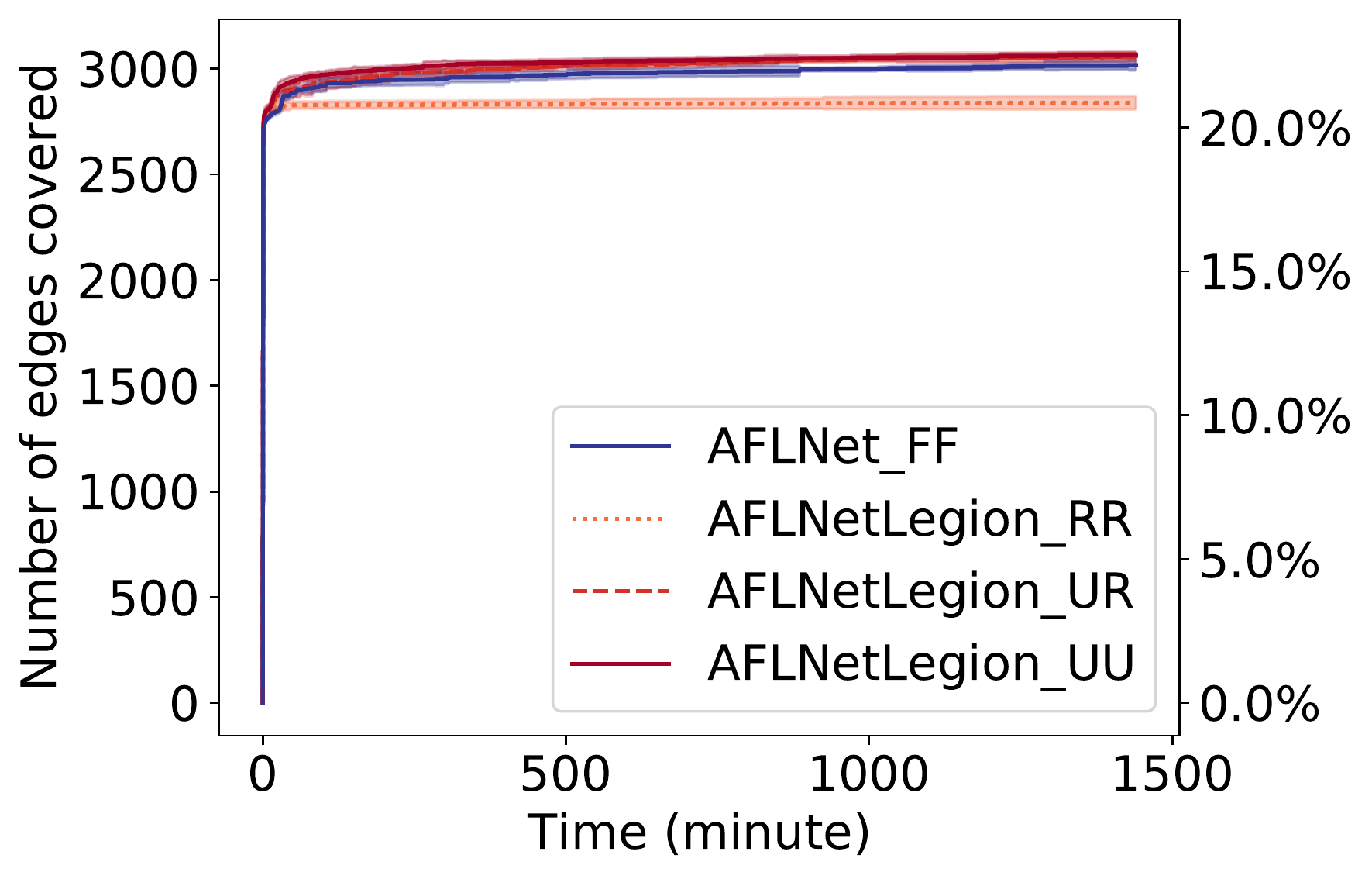}
        \caption{\texttt{Live555} branch coverage}
        \label{fig: overall live555 branch coverage}
    \end{subfigure}
    % \vskip\baselineskip
    % \begin{subfigure}[b]{0.32\textwidth}
    %     \centering
    %     \includegraphics[width=\textwidth]{Figures/openssh/l_favour_best.pdf}
    %     \caption{\texttt{OpenSSH} line coverage}
    %     \label{fig: overall openssh line coverage}
    % \end{subfigure}
    % \hfill
    % \begin{subfigure}[b]{0.32\textwidth}
    %     \centering
    %     \includegraphics[width=\textwidth]{Figures/openssl/l_random_best.pdf}
    %     \caption{\texttt{OpenSSL} line coverage}
    %     \label{fig: overall openssl line coverage}
    % \end{subfigure}
    % \hfill
    % \begin{subfigure}[b]{0.32\textwidth}
    %     \centering
    %     \includegraphics[width=\textwidth]{Figures/live555/l_favour_best.pdf}
    %     \caption{\texttt{Live555} line coverage}
    %     \label{fig: overall live555 line coverage}
    % \end{subfigure}
    \caption{\AFLNetLegion overall results.}
    \label{fig: aflnetlegion overall results}
\end{figure*}

Despite the promising results from the case study,
\AFLNetLegion failed to demonstrate largely better overall performance than its derivatives or the best performing algorithm from \AFLNet.
\cref{fig: aflnetlegion overall results}
repeats the experiment of \cref{fig: aflnet overall results}
with its best performing algorithm against
the three algorithms based on \AFLNetLegion.
% These three algorithms are
% \legionrr (uniform random state selection and seed selection)
% drawn in light red dotted lines,
% \legionur (UCT function for state selection and uniform random seed selection)
% in red dashed lines,
% \legionuu (UCT function for state selection and UCB for seed selection)
% in dark red solid lines.

In five out of six cases (except \texttt{OpenSSH}, discussed later),
\legionuu covered very similar amount of code
as the best-performing \AFLNet algorithm at the end.
For example,
the difference between their final branch coverage 
are proven to be statistically insignificant via t-tests.
in \texttt{LightFTP}, \texttt{OpenSSL}, and \texttt{Exim},
% the p-value from t-test on their final coverage is within $0.5\%$,
% and proving their performance to be statistically insignificant .
\legionuu outperformed the best-performing \AFLNet algorithm by less than $2.5\%$ branch coverage in \texttt{ProFTPD} and \texttt{Live555}, and largely under-performed it by $22\%$ in \texttt{OpenSSH}. 

In some cases,
three algorithms based on \AFLNetLegion can discover branches and lines faster at the beginning.
For example,
\legionuu and \legionur exhibits minor advantage at the first few minutes on
\texttt{LightFTP}, \texttt{ProFTPD}, and \texttt{OpenSSL}.
But at the rest time, all algorithms' performances are not visually distinguishable.

\legionrr is the worst performing algorithm in almost all cases as we expected.
Recall that the selection algorithm of \AFLNetLegion always
descends down the tree from the root and stops when a simulation child is selected.
With random selection,
state selection at each level of the tree becomes a Bernoulli process,
making the probability of selecting a simulation node at tree depth $n$
approximately governed by a binomial distribution,
where the likelihood of selecting that node is exponentially lower with a larger $n$.
For example, in \texttt{OpenSSH}, the number of times that a child node is selected for fuzzing is almost half of its parent.
% \MYTODO{Dongge: supporting statistics.}
The low probability of selecting deep tree nodes limits
\legionrr's ability to replay a long sequence of requests before mutation,
reducing its ability to explore the behaviour of deep server states.
% \Legion mitigated a similar issue via an optimisation that relies on symbolic execution, which is not available to \AFLNetLegion.

We also analysed why \AFLNetLegion's algorithms lose effectiveness on \texttt{OpenSSH}
with the statistics of $40$ experiments.
Firstly, it violates \AFLNetLegion assumption on the sufficiency of winning
(i.e., finding unique response code sequences).
The average probability of receiving a unique sequence from \texttt{OpenSSH} is
less than $1.2\%$ for \AFLNetLegion and approximately $2.5\%$ for \AFLNet.
As a comparison,
with the same volume of experiments ($40$ trials) of \texttt{ProFTPD},
the same probabilities are approximately $50\%$ and $31\%$ for \AFLNetLegion and \AFLNet respectively.
Recall that \AFLNetLegion's UCT algorithm evaluates a state with its
exploitation value (i.e., average unique sequence found by each selection) and
exploration value (i.e., inverse selection frequency compared with its siblings).
An extremely low winning rate will make the exploitation value negligible
and purely determines the potential of each state by how often it has not been selected,
which further lowers the winning probability.
The exploration value of the algorithm
optimistically estimates the potential of states within a confidence bound,
which will be ineffective if the winning rate is outside the bound.
This also made \AFLNetLegion's performance very unstable,
ranging from uncovering $6$ to $86$ unique response sequences on average of every $10$ trials.
Secondly,
MCTS's advantage in sequential decision making is useless in this particular benchmark program.
While this advantage helps \AFLNetLegion to select which requst seqeunce prefix to preserve,
most of the unique response sequences from \texttt{OpenSSH}
are found by not preserving any prefix.
For example,
FAVOUR found $89\%$ of the new sequences by selecting the first state for $13501.25$ times in an average trial.
% while \legionuu found almost all sequences from the tree root.
% On average,
% \aflnetff selected the first state $13501.25$ times;
% , about $60\%$ of them discovered a new sequence;
% \legionuu only selected the root state $8216.5$ times.
% at the discovery rate of $23\%$.

Another reason why \AFLNetLegion failed to demonstrate good performance like \Legion is due to the unavailability of symbolic execution.
Symbolic execution plays a crucial rule in optimising \Legion's performance in pruning nodes that have only one child or are not expected to find more execution paths,
but is not available to lightweight techniques like \AFLNetLegion.
% only relies on lightweight techniques.

%\input{ExperimentSetup}

%\input{ExperimentResult}

%\input{Threats}

%\input{RelatedWork}

\section{Discussion}
\label{sec: aflnetlegion - discussion}

By analyzing the results and other artefacts (e.g., fuzzing logs), we have identified two limitations of \AFLNet that could prevent the state selection algorithms from fully unlocking their potentials. Understanding these problems opens opportunities for future research in this interesting yet challenging topic of stateful network protocol fuzzing.

\iparagraph{Low fuzzing throughput (i.e. speed).} On average, \AFLNet and \AFLNetLegion achieved only 20 executions per second, which is hundreds of times lower than the normal throughput we could get while fuzzing stateless systems (e.g., media processing libraries). Stateful servers take longer time to reset and communications over the network is also much slower than file reading and writing operations. Having a low throughput in a given limited time budget, even if a state is correctly selected by a systematic algorithm like \AFLNetLegion, the fuzzer could not be able to generate enough test inputs to explore that state.

\iparagraph{State-aware and structure-aware mutations.} Servers under test normally expect well-structured inputs (i.e., the messages must adhere to some data format/input grammar). It means messages generated by random mutation operators currently supported by \AFLNet are likely rejected by the servers. Therefore, they may not go deep to explore more code paths. Structure-aware fuzzing approaches like AFLSmart \cite{aflsmart} and Nautilus \cite{nautilus} could be helpful. However, unlike stateless systems that require only one input model/grammar, stateful servers would require more---one for each state---because at each state, the expected input format could be different. Moreover, there could be dependencies across messages, making input generation for stateful servers more challenging.

\section{Conclusion}
\label{sec: aflnetlegion - conclusion}

In this paper, we have discussed the challenges in stateful network protocol fuzzing and the importance of state selection algorithms to explore the state space. We have also shared our investigation into the effectiveness and efficiency of six different state selection algorithms. After analyzing the experimental results and fuzzing artefacts, we have identified potential problems, such as low fuzzing throughput and low quality generated inputs, preventing state selection algorithms from fully unlocking their potentials. Our future plan is to find solutions for those hindering problems and extend our research to cover more algorithms.

\section{Acknowledgment}
This research was partially supported by the Commonwealth Scientific and Industrial Research Organisation’s Data61 under the Defence Science and Technology Group’s Next Generation Technologies Program. 
We also thank Google Cloud for giving us research credits to advance this work with their computing platform.

\iffalse
\textit{\color{blue} At most 10 pages for main text, 2 pages for references}
\textit{\color{blue} Use HotCRP format checker}
\fi

\clearpage

\bibliographystyle{IEEEtran}
\bibliography{AFLNetLegion}

\end{document}